\documentclass[prresearch,reprint,superscriptaddress]{revtex4-2}

\usepackage[version=3]{mhchem}
\usepackage{amsmath}
\usepackage{amssymb}
\usepackage{graphicx}
\usepackage{subfigure}
\usepackage{siunitx}
\usepackage[T1]{fontenc}
\usepackage{float}
\usepackage{mathptmx}
\usepackage{xspace}
\usepackage{enumitem}
\usepackage{tabularx}
\usepackage[frozencache,cachedir=.]{minted}
\usepackage{multirow}
\usepackage{booktabs}
\usepackage[hidelinks]{hyperref}
\usepackage{cleveref}
\usepackage{xr-hyper}
\usepackage[textsize=small]{todonotes}
\usepackage{filecontents}

\begin{document}

% File with author names and affiliations
\title{Optimal Intermediate Hamiltonians for Non-Equilibrium Free Energy Calculations:\\ A Numerical Study of Markov Models}

\author{David Beyer}
\email{david.beyer@mpinat.mpg.de}
\affiliation{Department of Theoretical and
Computational Biophysics, Max Planck Institute for Multidisciplinary Sciences, D-37077 Göttingen, Germany}

\author{Helmut Grubm\"{u}ller}
\email{hgrubmu@gwdg.de}
\affiliation{Department of Theoretical and
Computational Biophysics, Max Planck Institute for Multidisciplinary Sciences, D-37077 Göttingen, Germany}

\date{\today}

\begin{abstract}
The Jarzynski relation enables the estimation of equilibrium free energy differences from non-equilibrium, finite-time switching simulations.
These estimates usually converge poorly because rare trajectories dominate the exponential work average.
Here, we numerically determined and explored the sequence of intermediate Hamiltonians connecting initial and final states that minimize the mean squared error (MSE) of the Jarzynski estimator and thereby enhance convergence.
For discrete-time Markov models, an exact tilted-master-equation representation of the MSE in the large-sample limit, combined with automatic differentiation, enables efficient gradient-based minimization over all intermediate energies.
We applied our approach to three model systems of increasing complexity: a two-state model, a double-well potential, and a shifted potential well.
In all three systems, the optimal intermediate Hamiltonians jump at the initial and final times. 
Extensive Monte Carlo simulations show that optimal intermediates can reduce the MSE by more than an order of magnitude compared with linear and logarithmic interpolation, most strongly for large changes in the energy landscape. 
Remarkably, they need not dissipate less work than intermediates yielding larger errors.
Our results suggest heuristics for more efficient non-equilibrium free energy calculations of realistic molecular systems: optimal intermediate Hamiltonians jump at the initial and final times; for barrier-crossing problems, the barrier should be lowered rapidly and raised again later; and minimizing dissipation does not guarantee faster convergence.
\end{abstract}

\maketitle

\section{Introduction}

The concept of free energy plays a central role in the physics of soft and biological matter \cite{rubinstein2003polymer, phillips2012physical, doi2013soft, phillips2020molecular, frenkel2002understanding, muthukumar2023physics, Seifert_2025}. 
For example, free energy differences between thermodynamic states govern the binding of ligands to proteins \cite{berg2019biochemistry, phillips2020molecular}, the solvation of ions in water \cite{muthukumar2023physics}, and the phase behavior of macromolecules \cite{rubinstein2003polymer, hyman_annual_review}.
However, obtaining accurate estimates of free energy differences using molecular simulations remains challenging because present-day force fields
are of limited accuracy, and for large systems such as biomolecules, sampling is often insufficient.
Traditional approaches to calculating free energy differences, such as thermodynamic integration \cite{kirkwood_stat_mech} and free energy perturbation \cite{zwanzig1954high, widom1963some}, are based on equilibrium statistical mechanics and therefore often require long equilibration of all intermediate states.
In contrast, more recently developed non-equilibrium techniques like the Jarzynski relation  \cite{jarzynski1997nonequilibrium} and the Crooks fluctuation theorem \cite{crooks1999entropy} make it possible to calculate free energy differences between thermodynamic equilibrium states using work values obtained from non-equilibrium, finite-time switching simulations.
Here, we focus on the Jarzynski relation because the mean squared error of its unidirectional estimator can be approached analytically \cite{zuckerman2002theory,gore2003bias}.
The Jarzynski relation,
\begin{align}
    \left\langle e^{-W}\right\rangle = e^{-\Delta F},
    \label{eq:jarzynski}
\end{align}
relates the free energy difference $\Delta F \equiv F_{\mathrm{B}}-F_{\mathrm{A}}$ between two thermodynamic equilibrium states A and B to the non-equilibrium exponential work average $\left\langle e^{-W}\right\rangle$.
Note that we use reduced units with $\beta=1$ throughout the paper.
In \autoref{eq:jarzynski}, the average $\left\langle ...\right\rangle$ is taken over non-equilibrium trajectories evolving under a time-dependent Hamiltonian $H(t)$ that switches from an initial Hamiltonian, $H(0)=H_{\mathrm{A}}$, to a final Hamiltonian, $H(T)=H_{\mathrm{B}}$, in a finite time $T$.
The symbol $W$ denotes the work performed on the system along a trajectory.
For the Jarzynski relation to hold, the initial conditions for the non-equilibrium trajectories must be drawn from the equilibrium distribution of state A, $p_{\mathrm{A}}^{\mathrm{eq}}\propto \exp\left(- H_{\mathrm{A}}\right)$.
However, during the time evolution, the system can be driven far out of equilibrium and, in general, has not relaxed to $p_{\mathrm{B}}^{\mathrm{eq}}$ at the final time $T$.
\autoref{eq:jarzynski} is remarkably general and holds for different schemes of dynamical time evolution commonly used in molecular simulations, including Hamiltonian dynamics, Langevin dynamics, and discrete-time Monte Carlo (MC) evolution \cite{jarzynski1997nonequilibrium, jarzynski1997equilibrium, crooks1998nonequilibrium}.

The Jarzynski relation is an exact identity relating an ensemble average over non-equilibrium trajectories to the equilibrium free energy difference.
In molecular simulations, however, only a finite number $n$ of non-equilibrium trajectories can be sampled, each yielding a work value $W_i$.
For a finite number of work values, the free energy difference is approximated using the Jarzynski estimator $\Delta F^{(n)}$:
\begin{align}
        \Delta F \approx \Delta F^{(n)} \equiv -\ln\left(\frac{1}{n}\sum_{i=1}^{n}e^{-W_i}\right).
        \label{eq:jarzynski_estimator}
\end{align}
The Jarzynski estimator converges to the true free energy difference in the infinite-sample limit, $\lim_{n\rightarrow \infty}\Delta F^{(n)} = \Delta F$, but it is typically dominated by rare trajectories, leading to poor convergence \cite{lua2005practical, jarzynski2006rare, pohorille2010good}.
To improve the convergence of the Jarzynski estimator, various techniques have been proposed \cite{dellago2013computing}, including escorted trajectories \cite{vaikuntanathan2008escorted, Vaikuntanathan_escorted_jcp, lee2026estimating} and importance sampling of trajectories \cite{sun2003equilibrium, 10.1063/1.1760511, oberhofer2005biased, oberhofer2008optimum}.
The time dependence of the Hamiltonian offers another route for improvement: because the Jarzynski relation holds for \emph{arbitrary} $H(t)$ that satisfy the boundary conditions $H(0)=H_{\mathrm{A}}$ and $H(T)=H_{\mathrm{B}}$, it should be possible to accelerate the convergence of the Jarzynski estimator by optimally choosing the time-dependent sequence of intermediate Hamiltonians.

Determining optimal intermediate Hamiltonians for non-equilibrium problems is challenging because, even for simple systems, the space of all possible time-dependent Hamiltonians is extremely large.
Previous studies in stochastic thermodynamics \cite{seifert2012stochastic, peliti2021stochastic, Seifert_2025} that sought to minimize various objective functions through an appropriately designed time-dependent Hamiltonian have therefore mostly been restricted to a low-dimensional subspace of parametrized Hamiltonians.
In these approaches, a few external control parameters $\lambda(t)$ govern the time dependence of the Hamiltonian \cite{schmiedl2007optimal, schmiedl2009optimal, gingrich2016near, engel2023optimal, alvarado2026optimal}.
Schmiedl and Seifert analytically determined minimum-dissipation protocols for an overdamped colloidal particle in a harmonic trap \cite{schmiedl2007optimal}.
Dissipation is quantified by the mean dissipated work, $\left\langle W^{\mathrm{dis}}\right\rangle\equiv\left\langle W\right\rangle-\Delta F$, the mean work in excess of the free energy difference $\Delta F$ between the initial and final states.
Both for a moving trap and for a trap with a time-dependent spring constant, the minimum-dissipation protocols are discontinuous, i.e., the control parameter $\lambda(t)$ jumps at the initial and final times.
The existence of discontinuities in minimum-dissipation protocols was later also demonstrated for Brownian particles in anharmonic potentials \cite{then2008computing, geiger2010optimum, zhong2022limited, engel2023optimal}, as well as for underdamped \cite{gomez2008optimal} and active \cite{schuttler2025active, garcia2025optimal} particles, and is now recognized to be a general feature of such protocols \cite{blaber2021steps}.

Optimal protocols have also been studied in the context of non-equilibrium free energy calculations \cite{lindberg_optimizing, koning_optimizing, geiger2010optimum, cheng2025iterative}.
Geiger and Dellago \cite{geiger2010optimum} numerically determined protocols that minimize the mean squared error (MSE) between the true free energy difference and the Jarzynski estimator,
\begin{align}
        \mathrm{MSE} \equiv \mathbb{E}\left[\left(\Delta F^{(n)}-\Delta F\right)^2\right],
\label{eq:MSE}
\end{align}
where $\mathbb{E}\left[...\right]$ denotes the average over many realizations of $n$ non-equilibrium trajectories.
Studying an overdamped particle in a one-dimensional potential landscape, they found that minimum-error protocols generally differ from minimum-dissipation protocols, but, like them, show jumps at the initial and final times.
Moreover, their numerical results revealed that minimum-error protocols can dissipate even more work on average than naive linear protocols.

Although Geiger and Dellago made an important contribution to the foundations of non-equilibrium free energy calculations, their optimization was still restricted to a small subspace of the full space of Hamiltonians.
This raises the following question: Which intermediate Hamiltonians, unrestricted in their functional form, minimize the MSE for non-equilibrium free energy calculations? 
For equilibrium free energy perturbation, the analogous question has been answered: Reinhardt and Grubmüller determined unrestricted optimal intermediates variationally, assuming uncorrelated samples \cite{reinhardt2020determining}. 
This case is simpler, because the system is equilibrated in every intermediate sampling state. 
In contrast, for non-equilibrium calculations based on the Jarzynski relation, the system is equilibrated only in the initial state, and the free energy estimate depends on work values accumulated along entire trajectories. 
Presumably due to this complication, no comparable optimization has been reported for non-equilibrium free energy calculations.

Determining error-minimizing intermediate Hamiltonians is a formidable challenge, even for simple systems.
In this work, we therefore numerically determined optimal intermediates for different Markov models with a finite number of states evolving in discrete time.
For all studied systems, we found that optimal intermediate Hamiltonians jump at the initial and final times, that they can reduce the MSE by more than an order of magnitude compared to intermediates obtained from established interpolation schemes  such as linear interpolation of the Hamiltonians or of the Boltzmann factors, and that they need not dissipate less work than intermediates yielding a larger error.

The remainder of this article is structured as follows.
In \autoref{sec:model_methods}, we introduce the Markov models studied in this work (\autoref{sec:markov_model}) and explain how the Jarzynski relation is applied to their discrete-time evolution (\autoref{sec:path_weight}).
We then describe a tilted-master-equation approach for efficiently calculating optimal intermediates (\autoref{sec:minimizing}) and introduce the MC procedure used to benchmark the optimal intermediates (\autoref{sec:benchmark}).
In \autoref{sec:results}, we report and discuss numerical results for three model systems of increasing complexity: a two-state system (\autoref{sec:two_state}), a double-well potential (\autoref{sec:double_well}), and a shifted potential well (\autoref{sec:shifted_well}).
Finally, in \autoref{sec:outlook}, we summarize our findings and discuss perspectives for future work.

\section{Model and Methods}
\label{sec:model_methods}
\subsection{Discrete-Time Markov Model}
\label{sec:markov_model}

\begin{figure*}[ht]
\centering
\includegraphics[width=0.75\textwidth]{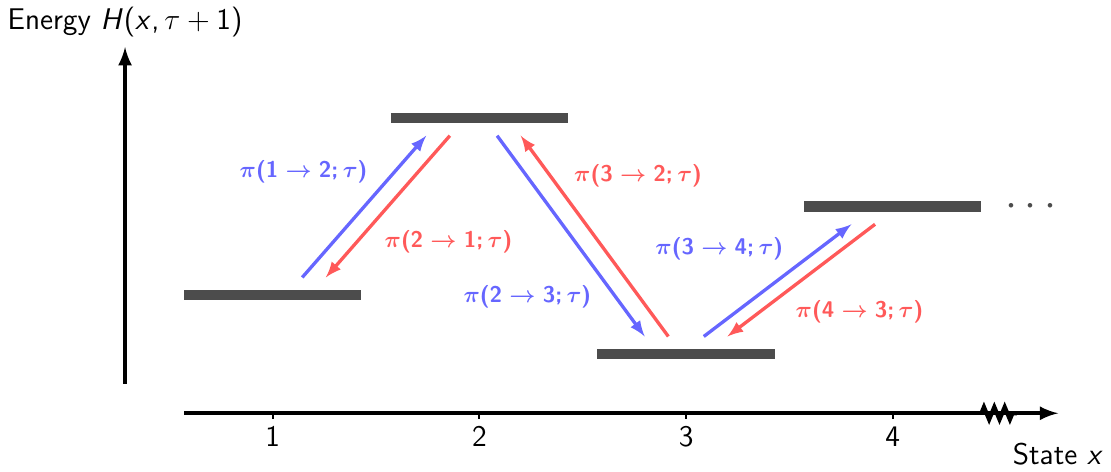}
\caption{\label{fig:schematic} Schematic representation of the discrete-time Markov models studied in this work.
The discrete states are arranged on a line, and transitions are allowed only between neighboring states.
Transition probabilities are given by the Glauber transition matrix elements $\pi(x\rightarrow y; \tau)$ (\autoref{eq:transition_matrix_glauber}), which depend on the time-dependent Hamiltonian $H(x,\tau+1)$.}
\end{figure*}

We investigated a series of simple model systems formulated as discrete-time Markov models with a finite state space. 
Besides enabling efficient numerical computation of optimal intermediates, this choice makes the systems toy models for Markov chain MC simulations, which are widely used in molecular modeling \cite{frenkel2002understanding, landau2021guide}.
\autoref{fig:schematic} shows the structure of our model systems: $m$ discrete states, labeled $x=1,2,...,m$, which are arranged on a line.
The populations $p(x,\tau)$ of the states evolve in a discrete MC time $\tau =0,1,2,...,N+1$ according to the master equation
\begin{align}
    p(x,\tau +1) = \sum_{y=1}^{m}\pi(y\rightarrow x;\tau)p(y,\tau).
    \label{eq:master_equation}
\end{align}
Here, $\pi$ is the transition matrix, which obeys detailed balance,
\begin{align}
    \frac{\pi\left(x\rightarrow y; \tau\right)}{\pi\left(y\rightarrow x; \tau\right)} = \exp\left[-\left(H(y,\tau+1)-H(x,\tau+1)\right)\right],
    \label{eq:detailed_balance}
\end{align}
and carries an explicit time dependence through the Hamiltonian $H(x,\tau)$.
The transition probabilities at time $\tau$ are governed by the
Hamiltonian at time  $\tau+1$ because the Hamiltonian is updated---thereby performing work on the system---\emph{before} the system jumps into a new state.
As discussed in the next subsection, this choice ensures that the Markov process is consistent with the Jarzynski relation \cite{jarzynski1997equilibrium, crooks1998nonequilibrium}.

Many different dynamical schemes satisfy \autoref{eq:detailed_balance};
we chose Glauber dynamics \cite{landau2021guide, glauber1963time} because the acceptance probability has a non-vanishing gradient for all finite energy differences, which makes it particularly suited to gradient-based minimization.
For Glauber dynamics, the transition matrix reads
\begin{widetext}
\begin{align}
\pi(x\rightarrow y; \tau) = 
\begin{cases}
g(x\rightarrow y)f\left(H(y,\tau+1)-H(x,\tau+1)\right), & \text{if } x \neq y,\\
1-\sum_{z\neq x}\pi(x\rightarrow z; \tau), & \text{if } x = y.
\end{cases}
\label{eq:transition_matrix_glauber}
\end{align}
\end{widetext}
In this definition, 
\begin{align}
    g(x\rightarrow y) \equiv \begin{cases}
\frac{1}{2}, & \text{if } y = x\pm 1 \\
0, & \text{otherwise}
\end{cases}
\label{eq:proposal_probability_glauber}
\end{align}
is the probability of proposing a new state $y$, which is non-zero only for neighboring states.
This choice determines the one-dimensional structure of the system, shown schematically in \autoref{fig:schematic}.
The function 
\begin{align}
    f(\Delta E)\equiv \frac{1}{1+\exp\left(\Delta E\right)}
    \label{eq:acceptance_probability_glauber}
\end{align}
is the Glauber acceptance probability \cite{landau2021guide, glauber1963time}.
\Cref{eq:master_equation,eq:transition_matrix_glauber,eq:proposal_probability_glauber,eq:acceptance_probability_glauber} define a thermodynamically consistent Markov chain for a driven system that can be treated either on the ensemble level, by solving the master equation, or on the trajectory level, by performing MC simulations with the specified transition probabilities.

So far, the physical nature of the Markov states has not been specified.
The schematic shown in \autoref{fig:schematic} suggests interpreting the Markov process as diffusion on a discretized potential landscape.
Introducing a lattice spacing $\delta x$ between neighboring states and a time increment $\delta \tau$, we expand the probabilities $p$ and transition matrix elements $\pi$ of the master equation (\cref{eq:master_equation}) in powers of $\delta x$ and $\delta \tau$.
The master equation then reduces to the Smoluchowski equation 
\begin{align}
    \partial_\tau p(x,\tau) = D\partial_x\left(\partial_x p(x,\tau)+p(x,\tau)\partial_x H(x,\tau)\right)
\end{align}
in the scaling limit $\delta x\rightarrow 0$, $\delta \tau\rightarrow 0$ with $\left(\delta x\right)^2=4 D\delta \tau$ kept fixed, where $D$ is the diffusion coefficient \footnote{Note that the factor of 4 in the equation $\left(\delta x\right)^2=4 D\delta \tau$ differs from the factor of 2 encountered in the scaling limit for a simple hopping process on a lattice. 
The additional factor of 2 appears because our MC scheme consists of two steps, a proposal step and an acceptance step.}.
Thus, the Markov models considered in the following can also be viewed as a discretized version of overdamped diffusive dynamics in a one-dimensional, time-dependent potential landscape $H(x,\tau)$.
This interpretation guided our choice of model systems (\autoref{sec:results}) and links them to biophysical applications such as barrier crossing and optical-trap experiments.

\subsection{Path Weight and Work along Stochastic Trajectories}
\label{sec:path_weight}
Jarzynski \cite{jarzynski1997equilibrium} and Crooks \cite{crooks1998nonequilibrium} showed that the discrete-time Markov dynamics introduced above satisfy the Jarzynski relation.
Consider the free energy difference $\Delta F = F_{N+1} - F_0$ between the initial Hamiltonian $H(x,0)$ and the final Hamiltonian $H(x,N+1)$.
Using the Markov process specified above with an arbitrary time-dependent Hamiltonian $H(x,\tau)$ that satisfies these boundary conditions, stochastic trajectories $(x_0,x_1,...,x_{N+1})$ can be generated.
If we draw the initial condition $x_0$ from the equilibrium distribution $p^{\mathrm{eq}}(x_0) = \exp(-H(x_0,0)) /Z_0$, where $Z_0$ is the partition function of the initial Hamiltonian, the path weight of a trajectory is given by 
\begin{align}
        \mathcal{P}(x_0,x_1,...,x_{N+1}) = p^{\mathrm{eq}}(x_0)\prod_{\tau=0}^{N}\pi\left(x_{\tau}\rightarrow x_{\tau+1};\tau\right).
        \label{eq:path_weight}
\end{align}
Evaluating the Jarzynski relation (\autoref{eq:jarzynski}) requires a consistent definition of the work performed along a trajectory \cite{jarzynski1997equilibrium, crooks1998nonequilibrium}.
Work is performed on the system when the Hamiltonian is updated between transitions. 
If the system is in state $x_\tau$ at time $\tau$, updating the Hamiltonian results in the work increment \cite{jarzynski1997equilibrium, crooks1998nonequilibrium}
\begin{align}
        \delta W(x_{\tau},\tau) = H(x_{\tau},\tau+1) - H(x_{\tau},\tau).
\end{align}
The work performed along a whole trajectory $(x_0,x_1,...,x_{N+1})$ is given by the sum over all work increments,
\begin{align}
        W(x_0,x_1,...,x_{N}) = \sum_{\tau=0}^{N}\delta W(x_{\tau},\tau).
        \label{eq:work_traj}
\end{align}
Note that the work does not depend on the final state $x_{N+1}$, because the Hamiltonian is updated to its final value before the system jumps into the final state.
Combining \autoref{eq:path_weight} and \autoref{eq:work_traj}, the exponential work average in the Jarzynski relation can be written as a path average:
\begin{align}
\begin{split}
    \left\langle e^{-W}\right\rangle &= \sum_{x_0}\sum_{x_1}...\sum_{x_{N}}\sum_{x_{N+1}} \mathcal{P}(x_0,x_1,...,x_{N},x_{N+1})e^{-W(x_0,x_1,...,x_{N})}\\
    &= \sum_{x_0}\sum_{x_1}...\sum_{x_{N}} \mathcal{P}(x_0,x_1,...,x_{N})e^{-W(x_0,x_1,...,x_{N})}.
\end{split}
\label{eq:path_average_jarzynski_markov}
\end{align}
Here, all sums $\sum_{x_i}$ run from $x_i=1$ to $x_i=m$.
For the second equality, we summed over the final state to simplify the equation: $\sum_{x_{N+1}}\pi(x_{N}\to x_{N+1};N)=1$.

\subsection{Minimizing the MSE}
\label{sec:minimizing}

Finding minimum-error intermediate Hamiltonians requires minimizing the MSE defined in \autoref{eq:MSE} with respect to $H(x,\tau)$.
In practice, this minimization is challenging because the search space is exceedingly large: for the Markov models introduced above, $H(x,\tau)$ comprises $N\cdot m$ independent energies.
A symmetry inherent in the Markov dynamics yields a
simplification: \autoref{eq:acceptance_probability_glauber} shows that the ``gauge transformation'' $H(x,\tau)\rightarrow H'(x,\tau) = H(x,\tau) + c(\tau)$ with an \emph{arbitrary} function $c(\tau)$ leaves the Markov dynamics invariant because the transition probabilities depend only on energy differences.
This transformation also leaves the MSE invariant, because $c(\tau)$ shifts all energy levels by the same amount and thus does not contribute to the fluctuating work. 
Therefore, $c(\tau)$ can be chosen arbitrarily (``gauge fixing'') and, for simplicity, the energy of state $x=1$ is always set to zero, leading to an optimization problem with $N\cdot(m-1)$ independent parameters.

Even with the gauge fixing in place, calculating the full MSE remains demanding.
Because $\Delta F^{(n)}$ depends on $n$ independent, random work values $W_i$, the MSE is an $n$-dimensional integral over the joint distribution of the work values, which is computationally expensive for realistic $n\approx 10^2$--$10^3$.
To further simplify the calculation, we followed the approach of Geiger and Dellago and considered the limit $n\gg 1$, for which the MSE reads \cite{zuckerman2002theory, gore2003bias, geiger2010optimum}
\begin{align}
        \mathrm{MSE}\sim \frac{1}{n}\left[\left\langle e^{-2(W-\Delta F)}\right\rangle-1\right], \quad \text{for }n \to \infty.
\end{align}
Because $\Delta F$ is fixed by the boundary Hamiltonians, minimizing this expression is equivalent to minimizing the proxy
\begin{align}
    \mathrm{MSE}'\equiv  \left\langle e^{-2W}\right\rangle.
\end{align}
For discrete-time Markov models, exponential work averages of the form $\left\langle e^{-\alpha W}\right\rangle$ ($\alpha\in\mathbb{R}$) can be efficiently calculated using a ``tilted'' master equation approach \cite{jarzynski1997equilibrium, touchette2009large}, which is implemented as follows.
To arrive at a computationally tractable form, using \autoref{eq:path_average_jarzynski_markov}, $\left\langle e^{-\alpha W}\right\rangle$ is expanded into a path average:
\begin{widetext}
\begin{align}
\begin{split}
        \left\langle e^{- \alpha W}\right\rangle =& \sum_{x_0}\sum_{x_1}...\sum_{x_{N}} \mathcal{P}(x_0,x_1,...,x_{N})e^{-\alpha W(x_0,x_1,...,x_{N})}\\
        =& \sum_{x_0}\sum_{x_1}...\sum_{x_{N}}\underbrace{p^{\mathrm{eq}}(x_0)e^{-\alpha\delta W(x_0,0)}}_{\equiv \tilde{p}(x_0;\alpha)}\times \underbrace{\pi\left(x_0\rightarrow x_1;0\right)e^{-\alpha\delta W(x_1,1)}}_{\equiv \tilde{\pi}(x_0\rightarrow x_1;0,\alpha)}\times ...\times \underbrace{\pi\left(x_{N-1}\rightarrow x_N;N-1\right)e^{-\alpha\delta W(x_N,N)}}_{\equiv \tilde{\pi}(x_{N-1}\rightarrow x_N;N-1,\alpha)}.
\end{split}
\end{align}
\end{widetext}
Here, we defined the tilted distribution vector $\tilde{\mathbf{p}}(\alpha)$ with components $\tilde{p}(x_0;\alpha)$ and tilted transition matrices $\tilde{\pi}(\tau,\alpha)$ with matrix elements $\tilde{\pi}(x_{\tau}\rightarrow x_{\tau+1};\tau,\alpha)$.
Using these definitions, the exponential work average is cast as a matrix-vector product,
\begin{align}
        \left\langle e^{- \alpha W}\right\rangle = \tilde{\mathbf{p}}(\alpha)^\mathsf{T}\tilde{\pi}(0,\alpha)\tilde{\pi}(1,\alpha) ...\tilde{\pi}(N-1,\alpha) \mathbf{1},
        \label{eq:matrix_product}
\end{align}
where $\mathbf{1}$ is the $m$-dimensional column vector consisting entirely of ones.
This representation yields $\mathrm{MSE}'$ \emph{exactly} (up to floating-point precision) at a computational cost of $\mathcal{O}\left(m^2N\right)$.

We implemented \autoref{eq:matrix_product} using the Python library JAX \cite{jax2018github}, which provides exact gradients via automatic differentiation.
Because $\left\langle e^{-\alpha W}\right\rangle$ reduces to the Jarzynski average for $\alpha=1$, the implementation was validated by comparing $-\ln \left\langle e^{-W}\right\rangle$ against the exactly known free energy difference for a series of randomly generated, time-dependent Hamiltonians.
To minimize $\mathrm{MSE}'$ with respect to the intermediate energies, we used the Adam optimizer \cite{kingma2014adam}, with a learning rate evolving according to the \texttt{warmup\_exponential\_decay\_schedule} implemented in Optax \cite{deepmind2020jax}. 
The learning rate schedule required to achieve convergence varied across systems and parameters, with the maximum learning rate ranging from approximately $10^{-1}$ to $10^2$.
The schedule comprised $10^4$ warmup steps, decay rates between $0.9$ and $0.95$, and $2000$--$5000$ transition steps.
The optimization was stopped once the norm of the gradient fell below $10^{-5}$ and the norm of the energy update fell below $10^{-10}$.
To assess the convergence of the results, for all systems discussed below, $\mathrm{MSE}'$ was minimized ten times, starting from different random initial guesses for the intermediate energies.
In all cases, the obtained intermediate energies differed by less than $10^{-3}$, indicating that the optimizations had converged.

\subsection{Benchmarking by Monte Carlo Simulations}
\label{sec:benchmark}

Because $\mathrm{MSE}'$ neglects the bias of the Jarzynski estimator and is valid only for $n\gg1$, we assessed the optimal intermediates using MC simulations, which yield the full MSE without this approximation. 
For every set of intermediates, we generated stochastic trajectories according to \Cref{eq:transition_matrix_glauber,eq:proposal_probability_glauber,eq:acceptance_probability_glauber}, with initial states drawn from $p^{\mathrm{eq}}(x_0)$ and the work accumulated along each trajectory via \autoref{eq:work_traj}. 
From $n$ such trajectories, one estimate $\Delta F^{(n)}$ was obtained via \autoref{eq:jarzynski_estimator}; we refer
to each such estimate as a replica. 
For each parameter set, we generated $10^5$ uncorrelated replicas for each of 18 approximately logarithmically spaced values $n \in \{1,\allowbreak 2,\allowbreak 4,\allowbreak 6,\allowbreak 8,\allowbreak 12,\allowbreak 18,\allowbreak 26,\allowbreak 37,\allowbreak 54,\allowbreak 78,\allowbreak 112,\allowbreak 162,\allowbreak 233,\allowbreak 335,\allowbreak 483,\allowbreak 695,\allowbreak 1000\}$, from which we estimated the MSE (\autoref{eq:MSE}) and its standard error \cite{janke02bfixed}.
The mean work $\langle W \rangle$ and its standard error were estimated using the work values from $10^8$ trajectories generated for each parameter set.

To benchmark the optimal intermediates, we repeated the MC simulations for different choices of $H(x,\tau)$ given by two established interpolation schemes.
The most widely used interpolation scheme we considered---for both equilibrium and non-equilibrium free energy calculations---is linear interpolation between the initial and final Hamiltonians,
\begin{equation}
  H(x,\tau) = \frac{\tau}{N+1}H(x,N+1) + \left(1-\frac{\tau}{N+1}\right)H(x,0).
  \label{eq:linear_interpolation}
\end{equation}
The second interpolation scheme instead interpolates the generalized Boltzmann factors $e^{-\alpha (H-F)}$ linearly,
\begin{widetext}
\begin{align}
   e^{-\alpha (H(x,\tau)-F_{\tau})} = \frac{\tau}{N+1}e^{-\alpha (H(x,N+1)-F_{N+1})} + \left(1-\frac{\tau}{N+1}\right)e^{-\alpha (H(x,0)-F_0)}.
   \label{eq:logarithmic_interpolation}
\end{align}
Here $\alpha\in\mathbb{R}$ is a scaling factor and the Boltzmann distribution is recovered for $\alpha=1$.
Solving \autoref{eq:logarithmic_interpolation} for $H(x,\tau)$ yields the ``logarithmic interpolation''
\begin{align}
    H(x,\tau) = -\frac{1}{\alpha}\ln\left[\frac{\tau}{N+1}e^{-\alpha (H(x,N+1)-\Delta F)} + \left(1-\frac{\tau}{N+1}\right)e^{-\alpha H(x,0)}\right] + \frac{1}{\alpha}\ln\left[\frac{\tau}{N+1}e^{\alpha \Delta F} + \left(1-\frac{\tau}{N+1}\right)\right],
    \label{eq:logarithmically_averaged_hamiltonians}
\end{align}
\end{widetext}
where the second term fixes $H(x=1,\tau)=0$.
We performed MC simulations for two previously used values of $\alpha$:
$\alpha=1$ corresponds to the reference state of enveloping distribution sampling \cite{christ_enveloping, christ_multiple_enveloping}; $\alpha=2$ approximates the sequence of optimal intermediates for free energy perturbation \cite{reinhardt2020determining}.

\begin{figure*}[ht]
\centering
\includegraphics[width=0.9\textwidth]{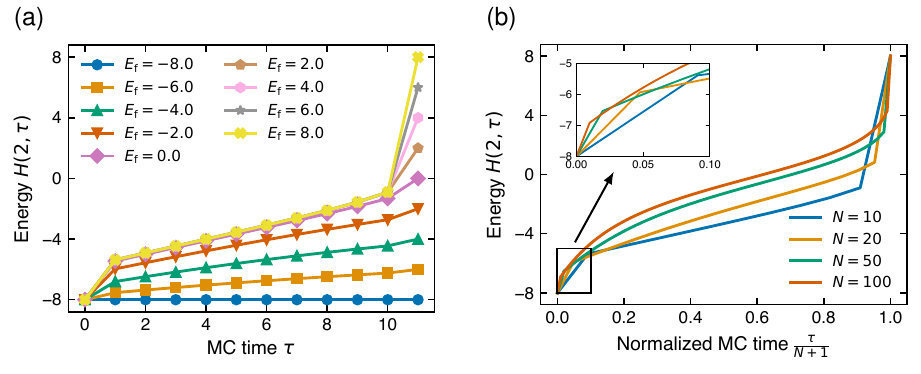}
\caption{\label{fig:protocols_two_state} Optimal intermediate Hamiltonians $H(2,\tau)$ for the two-state system with an initial energy of $E_{\mathrm{i}}=-8$.
(a): $H(2,\tau)$ for $N=10$ intermediates and different final energies $E_{\mathrm{f}}$.
(b): $H(2,\tau)$ for a final energy of $E_{\mathrm{f}}=8$ and different numbers of intermediates $N$. 
The MC time was normalized by the total number of steps $N+1$.
The inset magnifies the initial jump.}
\end{figure*}

\section{Results and Discussion}
\label{sec:results}
\subsection{Two-State Model}
\label{sec:two_state}
\subsubsection{Two-State Model with a Single Intermediate}
Before turning to numerical results, we consider a toy
model that can be treated analytically: a two-state system, $x\in\{1,2\}$, with a single intermediate Hamiltonian, $N=1$.
Using the gauge invariance (\autoref{sec:minimizing}), we set the energy of the first state to $H(1,\tau)=0$.
Furthermore, we set the initial energy of the second state to zero, $H(2,\tau=0)=0$, and the final energy to $H(2,\tau=2)\equiv E_{\mathrm{f}}$.
We seek the optimal value $\lambda^*$ of the intermediate energy $\lambda\equiv H(2,\tau=1)$.
For this toy system, $\mathrm{MSE}'$ reads
\begin{widetext}
\begin{align}
    \mathrm{MSE}'(\lambda) =& \frac{1}{2}\left[1-\frac{1}{2}f(\lambda )+\frac{1}{2}f(-\lambda)e^{-2\lambda } + \left(1-\frac{1}{2}f(-\lambda)\right)e^{-2E_{\mathrm{f}}} + \frac{1}{2}f(\lambda)e^{-2(E_{\mathrm{f}}-\lambda)}\right],
\end{align}
\end{widetext}
with the Glauber acceptance probability $f$ defined in \autoref{eq:acceptance_probability_glauber}.
To find the optimal intermediate energy $\lambda^*$, we set the derivative $\partial_{\lambda}\mathrm{MSE}'$ to zero, which yields
\begin{align}
    G(\lambda^*) = G(-\lambda^*)e^{-2E_{\mathrm{f}}}.
    \label{eq:nonlinear_equation_toy_model}
\end{align}
Here, we introduced the abbreviation $G(\lambda) \equiv f'(\lambda) + f'(-\lambda)e^{-2\lambda} + 2f(-\lambda)e^{-2\lambda}$.
\autoref{eq:nonlinear_equation_toy_model} has no closed-form solution.
However, in the limits $E_{\mathrm{f}}\rightarrow 0$ and $E_{\mathrm{f}}\rightarrow \pm \infty$, the equation can be solved approximately using standard methods of perturbation theory \cite{bender1999advanced}, yielding the asymptotic behavior (see Appendix \ref{sec:appendix_pert}):
\begin{align}
    \lambda^* \sim \begin{cases}
        \frac{E_{\mathrm{f}}}{2}, & \text{for }E_{\mathrm{f}} \to 0\\
        \pm\left[\sinh^{-1}(1) -\left(3 \sqrt{2} + 4\right)\cdot e^{\mp 2E_{\mathrm{f}}}\right], & \text{for } E_{\mathrm{f}} \to \pm\infty.
    \end{cases}
    \label{eq:asymptotics}
\end{align}
The solution shows two distinct regimes. 
If the final energy of state 2 is close to its initial value ($E_{\mathrm{f}} \to 0$), the optimal intermediate energy is $\lambda^*\sim E_{\mathrm{f}}/2$ and thus follows the linear interpolation.
In contrast, when the system is strongly driven out of equilibrium ($E_{\mathrm{f}} \to \pm\infty$), $\lambda^*$ saturates at $\pm\sinh^{-1}(1)\approx \pm 0.88$, which is approached exponentially.
Thus, the sequence of intermediates becomes distinctly asymmetric in time, consisting of a finite initial jump and a final jump that asymptotically grows linearly with $E_{\mathrm{f}}$.
As shown numerically in the next subsection, these features persist when multiple intermediate Hamiltonians are considered.

\autoref{eq:asymptotics} shows another interesting feature: when the final energy $E_{\mathrm{f}}$ changes sign, the optimal intermediate energy $\lambda^*$ does as well.
This feature is not restricted to the asymptotic limits but holds in general, as a consequence of the gauge invariance (\autoref{sec:minimizing}):
applying the transformation $c(\tau)=(0,-\lambda^*,-E_{\mathrm{f}})$ to the optimal Hamiltonian $H(1,\tau)=(0,0,0)$, $H(2,\tau)=(0,\lambda^*,E_{\mathrm{f}})$, we obtain $H'(1,\tau)=(0,-\lambda^*,-E_{\mathrm{f}})$, $H'(2,\tau)=(0,0,0)$.
After exchanging the labels of states 1 and 2, $H'$ is the ``mirrored'' version of the original time-dependent Hamiltonian $H$: $H'=-H$.
This symmetry extends to any number of intermediates and to arbitrary initial and final energies; we therefore used it as a consistency check for the numerical results presented in the next subsection.

\subsubsection{Two-State Model with Multiple Intermediates}

For more than a single intermediate, even the simple two-state model cannot be treated analytically.
Therefore, we resorted to the numerical minimization procedure described in \autoref{sec:minimizing} and determined optimal intermediate Hamiltonians $H(2,\tau)$ for different numbers of intermediates, $N$, different initial energies, $E_{\mathrm{i}}\equiv H(2,\tau=0)$, and different final energies, $E_{\mathrm{f}}\equiv H(2,\tau=N+1)$.
Specifically, we computed the optimal intermediate Hamiltonians for all combinations of $N\in\{10,20,50,100\}$, $E_{\mathrm{i}}\in\{-8,-6,-4,-2,0,2,4,6,8\}$, and $E_{\mathrm{f}}\in\{-8,-6,-4,-2,0,2,4,6,8\}$.
As a consistency check, we verified that for every combination $(N,E_{\mathrm{i}},E_{\mathrm{f}})$, the mirrored intermediates $-H(2,\tau)$ agreed within $10^{-3}$ with those computed for $(N,-E_{\mathrm{i}},-E_{\mathrm{f}})$.
All 324 numerically determined curves are provided in the Supplemental Material (\autoref{esi:fig:exhaustive_protocols_two_state_1} and \autoref{esi:fig:exhaustive_protocols_two_state_2}).
In the following, we focus on a representative subset of these results to discuss the trends that hold across the whole data set.

We first consider the influence of the change in energy, $\Delta E = E_{\mathrm{f}}-E_{\mathrm{i}}$, on optimal intermediate Hamiltonians for a fixed number of intermediates $N$.
\autoref{fig:protocols_two_state}(a) shows the optimal time dependence of $H(2,\tau)$ for $N=10$, $E_{\mathrm{i}}=-8$ and different values of $E_{\mathrm{f}}$.
For $\Delta E=0$, $H(2,\tau)$ is constant.
For $\Delta E=2$---a relatively small change in energy---the optimal intermediate energies increase approximately linearly with $\tau$. 
As $\Delta E$ increases further, the function $H(2,\tau)$ develops jumps at the initial and final times, with an approximately linear increase in between.
The initial jump develops first and saturates at $H(2,\tau=1)-H(2,\tau=0)\approx 2.6$ for $\Delta E \gtrsim 8$.
The final jump, from $\tau=10$ to $\tau=11$, continues to grow with $\Delta E$, whereas the remaining intermediate energies become independent of $\Delta E$.
This behavior is qualitatively similar to the result obtained for the analytical toy model in the previous section, where the initial jump $H(2,1)-H(2,0)$ saturates, whereas the final jump $H(2,2)-H(2,1)$ grows without bound.
Moreover, the jumps in the optimal $H(2,\tau)$ resemble those of work- and error-minimizing protocols previously obtained for colloidal particles in various potential landscapes \cite{schmiedl2007optimal, then2008computing, geiger2010optimum}.
For work-minimizing protocols, jumps arise at the boundaries because they enable slower driving in between, which reduces dissipation.
In contrast, for the error-minimizing protocols, the reason for the jumps is less obvious and will be discussed below.

Next, we investigated how the optimal intermediate Hamiltonians depend on the rate of switching, which is set by the number $N$ of intermediates.
\autoref{fig:protocols_two_state}(b) compares the optimal intermediates $H(2,\tau)$ for $E_{\mathrm{i}}=-8$, $E_{\mathrm{f}}=8$, and different values of $N$.
To facilitate a direct comparison of the different curves, the MC time was normalized by the total number of steps $N+1$.
The figure shows that, for all values of $N$, the optimal $H(2,\tau)$ contains jumps at the initial and final times, the final jump being much more pronounced than the initial one.
Comparing the curves for different values of $N$, the magnitudes of both jumps decrease monotonically with $N$.
Again, this behavior resembles that of optimal protocols for colloidal particles, whose jumps also shrink with increasing switching time \cite{schmiedl2007optimal, then2008computing, geiger2010optimum}.

\begin{figure}[ht]
\centering
\includegraphics[width=0.79\columnwidth]{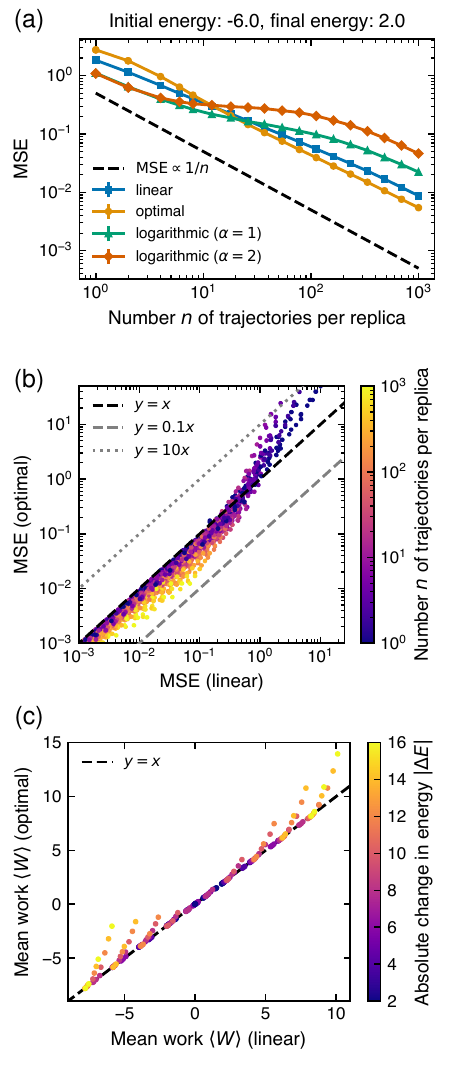}
\caption{\label{fig:mse_two_state} 
Performance of optimal intermediates for a two-state system.
(a): Dependence of the MSE for linear, logarithmic, and optimal intermediates on the number $n$ of trajectories per replica. 
The initial energy is set to $E_{\mathrm{i}}=-6$, the final energy is set to $E_{\mathrm{f}}=2$, and the number of intermediate steps is $N=10$.
The black dashed line shows the expected power-law decay $\mathrm{MSE}\propto 1/n$ for $n\rightarrow \infty$.
(b): Correlation plot of the MSE for linear and optimal intermediates.
(c): Correlation plot of the mean work performed on the system for linear and optimal intermediates. 
In all subfigures (a)--(c), error bars corresponding to the standard error of the mean of the shown quantities are included, but smaller than the symbol size.}
\end{figure}

Next, we asked whether---and to what extent---the optimal intermediates actually reduce the true MSE, including the bias neglected by $\mathrm{MSE}'$.
As an example, \autoref{fig:mse_two_state}(a) shows the MSE for linear, logarithmic, and optimal intermediates as a function of the number $n$ of trajectories per replica for $E_{\mathrm{i}}=-6$, $E_{\mathrm{f}}=2$, and $N=10$, calculated using MC simulations (\autoref{sec:benchmark}).
For all cases, the MSE decays with increasing $n$, approaching the power law $\mathrm{MSE}\propto 1/n$ expected for $n\rightarrow\infty$.
Three regimes can be distinguished.
For $n\lesssim 10$, the MSE obtained for linear and logarithmic intermediates is smaller than that for optimal intermediates; for $10\lesssim n\lesssim 30$, the MSE for optimal intermediates crosses that for linear and logarithmic intermediates, and for $n\gtrsim 30$, the optimal intermediates systematically outperform both interpolation schemes, with an MSE up to almost an order of magnitude smaller for large $n$.
The poor performance of the optimal intermediates at small $n$ stems directly from our use of the proxy $\mathrm{MSE}'$: for small $n$, the Jarzynski estimator is strongly biased \cite{zuckerman2002theory, gore2003bias, geiger2010optimum}, and for this system, the bias dominates the $\mathrm{MSE}$ but is neglected by $\mathrm{MSE}'$.
As $n$ increases, the contribution of the bias to the $\mathrm{MSE}$ becomes less important, and $\mathrm{MSE}'$ becomes a better approximation, eventually leading to optimal intermediates that outperform the linear and logarithmic ones.

To compare the performance of linear and optimal intermediates across the whole parameter range $(N, E_{\mathrm{i}}, E_{\mathrm{f}}, n)$, \autoref{fig:mse_two_state}(b) (see also \autoref{esi:fig:correlation_plots_mse_two_state_esi}) shows the MSE for optimal intermediates against that for linear intermediates.
Consistent with \autoref{fig:mse_two_state}(a), the optimal intermediates yield less accurate free energy estimates than the linear ones for small $n$, with an MSE that is larger by an order of magnitude in extreme cases.
For $n\gtrsim 100$, the optimal intermediates consistently outperform the linear ones.
However, the optimal intermediates perform markedly better only for $10^{-3}\lesssim \mathrm{MSE}\lesssim 5\cdot 10^{-1}$ (MSE of the linear intermediates), whereas for $\mathrm{MSE}\lesssim 10^{-3}$, both perform similarly (\autoref{esi:fig:correlation_plots_mse_two_state_esi}(a)).
Comparing simulations across different values of $\Delta E$ (\autoref{esi:fig:correlation_plots_mse_two_state_esi}(c) and (d)) and for different numbers of intermediates $N$ (\autoref{esi:fig:correlation_plots_mse_two_state_esi}(e) and (f)), the improvement from optimal intermediates was largest for large $\Delta E$ and small $N$.
This observation is expected, as for large $\Delta E$, linear and optimal intermediates differ the most.
Moreover, for large $\Delta E$ and small $N$, the system is strongly driven out of equilibrium, potentially leaving greater room for improvement through optimally designed intermediates.
Besides the linear intermediates, we considered logarithmic intermediates with $\alpha=1$ and $\alpha=2$.
Results for these MC simulations are reported in the Supplemental Material (\autoref{esi:fig:correlation_plots_mse_two_state_log_linear_esi} and \autoref{esi:fig:correlation_plots_mse_two_state_avi_esi}) and follow the same trends as observed in the linear case.
Surprisingly, logarithmic intermediates performed worse than linear ones, although $\alpha=2$ approximates the optimal sequence of intermediates for free energy perturbation \cite{reinhardt2020determining}.
This finding shows that the optimal intermediates for equilibrium-based free energy calculations do not transfer to the non-equilibrium case.

Previous works in stochastic thermodynamics conjectured that protocols with a small amount of dissipated work should also improve non-equilibrium free energy estimates \cite{schmiedl2007optimal, engel2023optimal}.
However, whereas work- and error-minimizing protocols coincide for shifted harmonic traps, Geiger and Dellago numerically showed that error-minimizing optimal protocols do not necessarily minimize the dissipated work, and may even dissipate more work than a naive linear protocol \cite{geiger2010optimum}.
To study the dissipation of work for our two-state system, we calculated the mean work for optimal, linear, and logarithmic intermediates using the MC simulations. 
\autoref{fig:mse_two_state}(c) shows a correlation plot of the mean work for optimal and linear intermediates, which reveals that the mean work for optimal intermediates is consistently larger than that for linear intermediates.
The observed difference in mean work was particularly pronounced for large values of $\Delta E$ and fast switching (see \autoref{esi:fig:correlation_plots_work_two_state_esi}(b))---exactly those cases in which the optimal intermediates achieve the largest improvement over the linear ones.
For logarithmic intermediates with $\alpha=1$, we observed the same qualitative behavior as in the linear case (\autoref{esi:fig:correlation_plots_work_two_state_log_linear_esi}).
Logarithmic intermediates with $\alpha=2$ also dissipated more work than optimal intermediates over large ranges of the studied parameter space, but slightly less for slow switching (\autoref{esi:fig:correlation_plots_work_two_state_avi_esi}). 
Taken together, these findings emphasize that there is no simple correspondence between work- and error-minimizing intermediates, and that therefore the dissipated work is not necessarily a good proxy for the mean error of non-equilibrium free energy calculations.

\subsection{Double-Well Potential}
\label{sec:double_well}

\begin{figure*}[ht!]
\centering
\includegraphics[width=\textwidth]{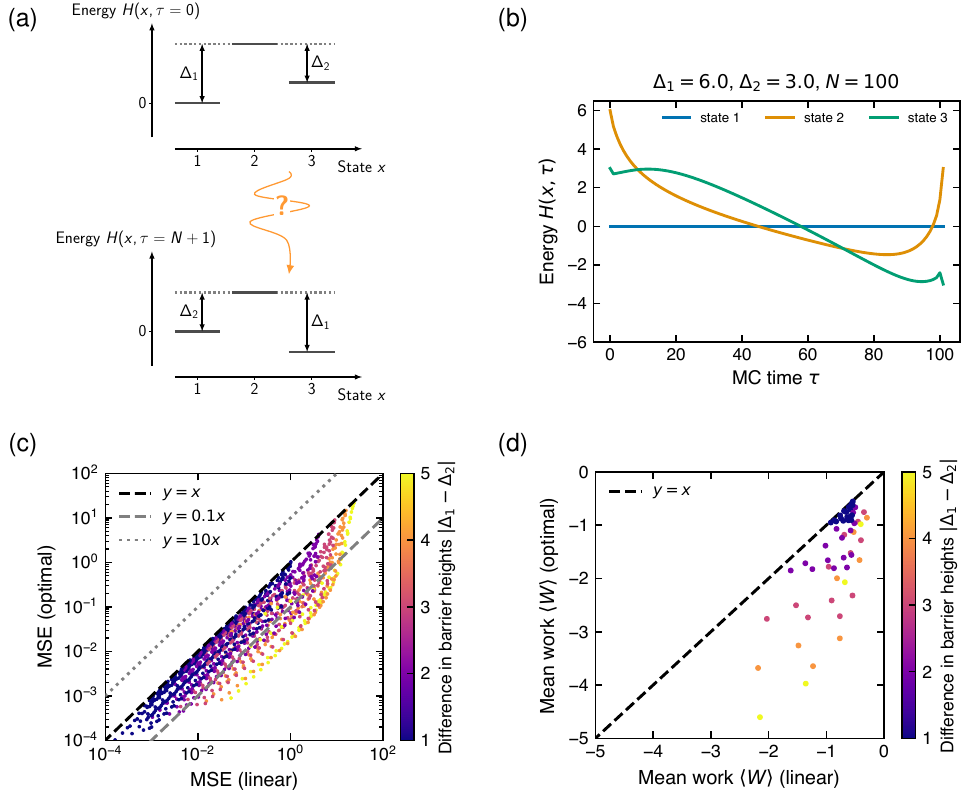}
\caption{\label{fig:main_figure_three_state} 
Optimal intermediates and their performance for the double-well potential.
(a): Toy model for a double-well potential with barrier heights $\Delta_1$ and $\Delta_2$.
(b): Optimal intermediate Hamiltonian for a double-well potential with $\Delta_1=6$, $\Delta_2=3$, and $N=100$ intermediate MC steps.
(c): Correlation plot of the MSE for linear and optimal intermediates.
(d): Correlation plot of the mean work performed on the system for linear and optimal intermediates. 
For subfigures (c) and (d), error bars corresponding to the standard error of the mean of the shown quantities are included, but smaller than the symbol size.
}
\end{figure*}

As a second example, we investigated a three-state system serving as a toy model of a double-well potential (see \autoref{fig:main_figure_three_state}(a)).
At $\tau=0$, state $x=1$ is the global energy minimum (barrier height $\Delta_1$) and state $x=3$ a local minimum (barrier height $\Delta _2$); at $\tau=N+1$, the two minima have exchanged roles.
The middle state, $x=2$, acts as a transition state.
This setup mimics barrier-crossing problems encountered, for example, when determining the unfolding free energy of a nucleic acid molecule in an optical trap \cite{engel2023optimal}.
We determined the optimal intermediates for all combinations of $N\in\{10,20,50,100\}$, $\Delta_1\in\{2,3,4,5,6\}$, and $\Delta_2 \in\{1,2,3,4,5\}$ with $\Delta_2 < \Delta _1$, amounting to 60 cases.
Again, all numerically determined optimal intermediates are provided in the Supplemental Material (\Crefrange{esi:fig:exhaustive_protocols_three_state_1}{esi:fig:exhaustive_protocols_three_state_3}).

\autoref{fig:main_figure_three_state}(b) shows the optimal intermediates for a representative example ($\Delta_1=6$, $\Delta_2=3$, and $N=100$).
Compared with the two-state system, the optimal intermediate energies have a more complex time dependence.
Specifically, the energy $H(2,\tau)$ of the transition state (state 2) initially drops sharply and then continues to decrease, eventually falling below the energy of state 1.
At $\tau\approx 90$, it begins to increase again and eventually jumps to its final value.
For all 60 parameter combinations, $H(2,\tau)$ varies non-monotonically (\autoref{esi:fig:exhaustive_protocols_three_state_1}--\autoref{esi:fig:exhaustive_protocols_three_state_3}); this feature likely accelerates the transition from state 1 to state 3.
The energy of state 3 also varies non-monotonically and shows small jumps at the initial and final times.
Notably, the energies of states 2 and 3 cross twice during the time evolution.

Next, we performed MC simulations to compare the performance of linear, logarithmic, and optimal intermediates for the double-well potential.
\autoref{fig:main_figure_three_state}(c) shows a correlation plot of the MSE obtained for linear and optimal intermediates; corresponding plots for the logarithmic intermediates are provided in the Supplemental Material (\autoref{esi:fig:correlation_plots_mse_three_state_log_linear_esi} and \autoref{esi:fig:correlation_plots_mse_three_state_avi_esi}).
In contrast to the two-state system, the MSE for optimal intermediates was lower than that for linear and logarithmic intermediates in all cases studied.
This difference suggests that the bias of the Jarzynski estimator is less pronounced for the double-well potential, leading to better performance of optimal intermediates, even for small values of $n$.
Moreover, \autoref{fig:main_figure_three_state}(c), \autoref{esi:fig:correlation_plots_mse_three_state_log_linear_esi}, and \autoref{esi:fig:correlation_plots_mse_three_state_avi_esi} show that, for the double-well potential, the optimal intermediates frequently reduce the MSE by more than an order of magnitude.
The largest improvement of optimal over linear intermediates was achieved for large differences in barrier heights, $|\Delta _1-\Delta _2|$, that is, for large changes in the energy landscape between the initial and final Hamiltonians.
This finding agrees qualitatively with the results for the two-state system.
Across different numbers of intermediates $N$ (\autoref{esi:fig:correlation_plots_mse_three_state_esi}(b), \autoref{esi:fig:correlation_plots_mse_three_state_log_linear_esi}(c), and \autoref{esi:fig:correlation_plots_mse_three_state_avi_esi}(c)), the improvement was largest for large $N$, in marked contrast to the two-state system.

Lastly, we compared the mean work for linear, logarithmic, and optimal intermediates (\autoref{fig:main_figure_three_state}(d), \autoref{esi:fig:correlation_plots_work_three_state_log_linear_esi}, and \autoref{esi:fig:correlation_plots_work_three_state_avi_esi}).
Here, the optimal intermediates consistently yielded a smaller mean work.
This difference was most pronounced for large values of $|\Delta _1-\Delta _2|$ and slow driving, precisely those cases in which the largest improvement in the MSE was observed.
In summary, for the double-well potential the optimized intermediates lowered the barrier within the first few steps, enabling rapid barrier crossing, and restored it only in the last ones.

\subsection{Shifted Potential Well}
\label{sec:shifted_well}

\begin{figure*}[p!]
\centering
\includegraphics[width=\textwidth]{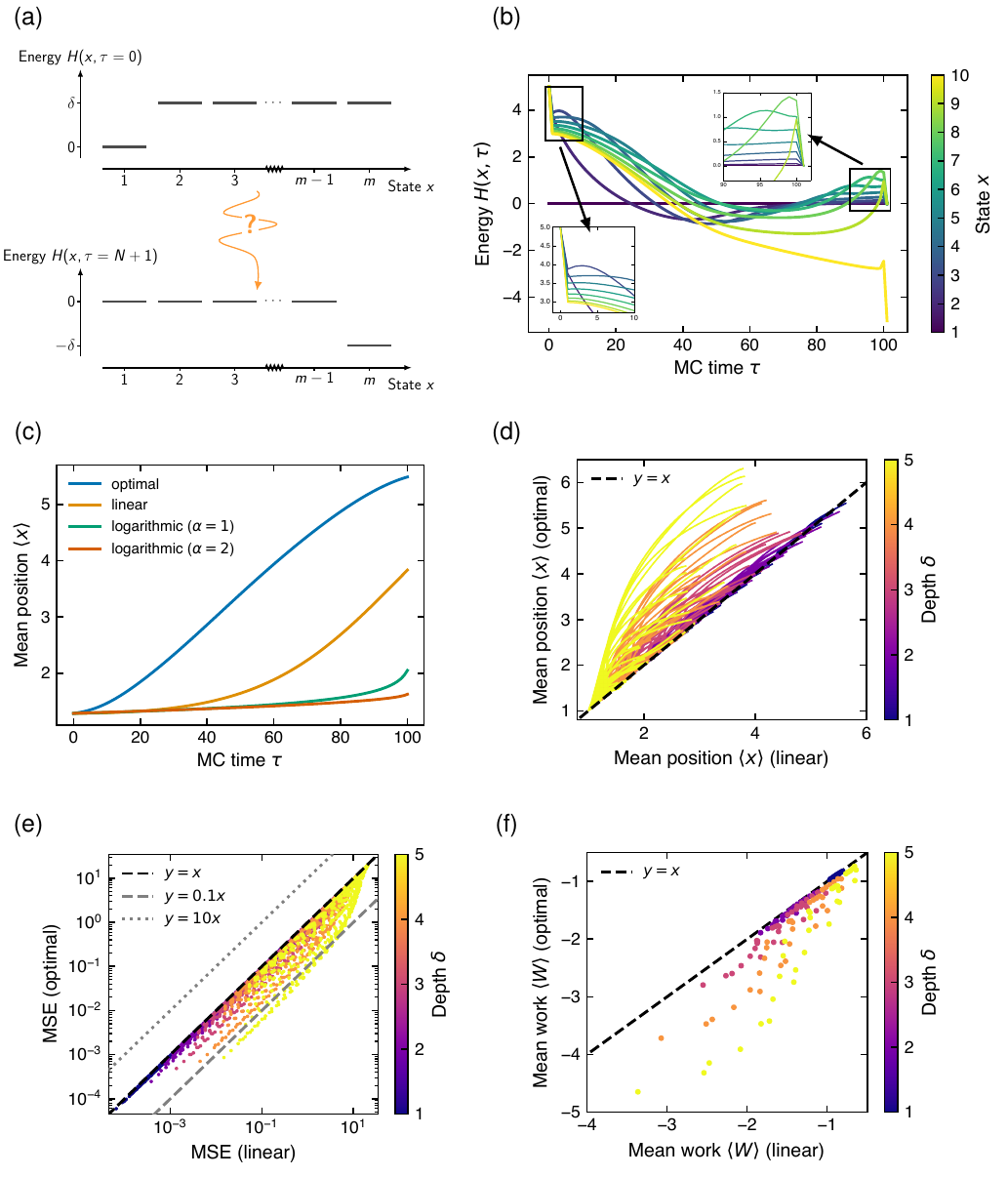}
\caption{\label{fig:main_figure_single_well} 
Optimal intermediates and their performance for the shifted potential well.
(a): Toy model for a potential well of depth $\delta$ that is shifted from $x=1$ at time $\tau=0$ to $x=m$ at time $\tau=N+1$.
(b): Optimal intermediate Hamiltonian for a shifted potential well of depth $\delta=5$, with $m=10$ states and $N=100$ intermediate MC steps.
The insets show magnified views of the initial and final jumps.
(c): Mean position $\left\langle x\right\rangle$ as a function of $\tau$ for $\delta=5$, $m=10$ states, and $N=100$, shown for optimal, linear, and logarithmic ($\alpha=1$ and $\alpha=2$) intermediates.
(d): Correlation plot of the time-dependent mean position $\left\langle x\right\rangle$ obtained for optimal and linear intermediates.
The curves are colored according to the depth $\delta$ of the potential well.
(e): Correlation plot of the MSE for linear and optimal intermediates.
(f): Correlation plot of the mean work performed on the system for linear and optimal intermediates. 
For subfigures (e) and (f), error bars corresponding to the standard error of the mean of the shown quantities are included, but smaller than the symbol size.
}
\end{figure*}

Finally, we determined optimal intermediate Hamiltonians for a system with more states (\autoref{fig:main_figure_single_well}(a)), a potential well of depth $\delta$, located at $x=1$ at time $\tau=0$ and shifted to $x=m$ at time $\tau=N+1$.
This system is a toy model of the optical-trap setups studied in previous works \cite{schmiedl2007optimal,then2008computing,geiger2010optimum}, where the trap center was shifted in finite time.
The crucial difference from these studies is that our approach does not impose any shape of the potential landscape on the intermediate Hamiltonians.
We determined the optimal intermediates for all combinations of $N\in\{10,20,50,100\}$, $m\in\{3,4,5,6,7,8,9,10\}$, and $\delta \in\{1,2,3,4,5\}$, amounting to 160 cases.
All numerically determined optimal intermediates are provided in the Supplemental Material (\Crefrange{esi:fig:exhaustive_protocols_pulling_single_well_1}{esi:fig:exhaustive_protocols_pulling_single_well_8}).

\autoref{fig:main_figure_single_well}(b) shows the optimal intermediates for a representative example ($\delta=5$, $m=10$, and $N=100$).
As for the previous systems, the optimal intermediates show jumps at the initial and final times.
The magnitude of these jumps generally increases with the distance from the initial position of the well.
Moreover, we found that the magnitude of the jumps generally decreases with the number $N$ of intermediates (\Crefrange{esi:fig:exhaustive_protocols_pulling_single_well_1}{esi:fig:exhaustive_protocols_pulling_single_well_8}).
Between the jumps, the energy levels vary non-monotonically and occasionally cross.

The complex behavior of the optimal intermediates renders a straightforward interpretation challenging.
Intuitively, one would expect the optimal intermediates to transport the probability distribution from its initial peak at $x = 1$ to its final peak at $x = m$ more efficiently than both linear and logarithmic intermediates.
To test this expectation, we numerically solved the master equation (\autoref{eq:master_equation}) for linear, logarithmic, and optimal intermediates to obtain the time-dependent probability distributions $p(x,\tau)$.
For $\delta=5$, $m=10$, and $N=100$, \autoref{fig:main_figure_single_well}(c) shows the resulting mean position $\left\langle x\right\rangle(\tau) = \sum_{x=1}^{m}xp(x,\tau)$.
As can be seen, the optimal intermediates transport the system at almost constant speed across the time-dependent potential landscape, leading to a final value of $\left\langle x\right\rangle(\tau=100)\approx 5.5$.
In contrast, for linear intermediates, the system remains near $x = 1$ for a much longer time but accelerates as $\tau$ increases, leading to a final position of $\left\langle x\right\rangle(\tau=100)\approx 4$.
Logarithmic intermediates exhibit a similar accelerating behavior, albeit at a lower overall rate, resulting in even less advanced final positions.

The acceleration observed for linear and logarithmic intermediates can be explained as follows. 
Because the energy levels near the initial well change only gradually, the system initially remains at $x=1$. 
Once the energy difference between states 1 and 2 becomes sufficiently small for the system to leave state 1 with appreciable probability, the probability for the system to diffuse freely on the almost flat part of the energy landscape between $x=2$ and $x=m-1$ increases, leading to a rapid increase of $\left\langle x\right\rangle$.
To compare the dynamics under linear, logarithmic, and optimal intermediates for all parameter combinations, we plotted the time-dependent mean position $\left\langle x\right\rangle(\tau)$ of the optimal case against that of the linear and logarithmic cases, respectively (\autoref{fig:main_figure_single_well}(d), \autoref{esi:fig:correlation_plots_mean_position_log_linear_esi}, and \autoref{esi:fig:correlation_plots_mean_position_avi_esi}).
In most cases, the optimal intermediates transport the probability distribution farther than the linear and logarithmic intermediates.
The difference is most pronounced for deep potential wells (large $\delta$), because, in the linear and logarithmic cases, the particle remains trapped at $x=1$ for a long time, whereas the optimal intermediates enable more efficient transport.

Lastly, we performed MC simulations to quantitatively assess the performance of the error-minimizing optimal intermediates. 
\autoref{fig:main_figure_single_well}(e) correlates the MSE values obtained for linear and optimal intermediates; corresponding plots for logarithmic intermediates are provided in the Supplemental Material (\autoref{esi:fig:correlation_plots_mse_single_well_log_linear_esi} and \autoref{esi:fig:correlation_plots_mse_single_well_avi_esi}).
In contrast to the two-state system and in agreement with the double-well potential, the MSE obtained for optimal intermediates was always smaller than that for linear and logarithmic intermediates, even for small $n$ (see \autoref{esi:fig:correlation_plots_mse_single_well_esi}, \autoref{esi:fig:correlation_plots_mse_single_well_log_linear_esi}, and \autoref{esi:fig:correlation_plots_mse_single_well_avi_esi}).
Comparing the MSE across different parameters, we found that the reduction in the MSE was largest for deep potential wells (large $\delta$) and slow driving (large $N$), similar to the results for the double-well potential.
Correlating the mean work for linear, logarithmic, and optimal intermediates (\autoref{fig:main_figure_single_well}(f), \autoref{esi:fig:correlation_plots_work_single_well_log_linear_esi}, and \autoref{esi:fig:correlation_plots_work_single_well_avi_esi}), we found that the optimal intermediates consistently yielded a smaller mean work.
As for the MSE, this difference was most pronounced for deep potential wells and slow driving.
In summary, for the shifted potential well the optimized intermediates moved the probability distribution at almost constant speed, whereas for linear and logarithmic intermediates it stayed near the initial position and then accelerated, resulting in a less advanced final position.

\section{Summary and Outlook}
\label{sec:outlook}

We have numerically determined optimal intermediate Hamiltonians that minimize the mean squared error of the Jarzynski estimator for simple Markov models.
To calculate the optimal intermediates, we combined the tilted master equation formalism with automatic differentiation, enabling rapid gradient-based optimization.
For all studied systems, we also performed extensive MC simulations to compare the optimal intermediates with linear and logarithmic interpolation.

As a first model, we considered a simple two-state system. 
Treating the case of a single intermediate perturbatively, we found that the optimal intermediate energy follows the linear interpolation for small energy differences, but saturates at a finite value for large energy differences.
Turning to numerical results for multiple intermediates, we found that the optimal intermediates show marked, asymmetric jumps at the initial and final times, with the initial jump saturating for large energy differences, whereas the final jump grows without bound.
Moreover, we observed that the magnitude of the jumps generally decreases with the number of intermediates, that is, with slower driving.
Comparing linear, logarithmic, and optimal intermediates using MC simulations, we found that the optimal intermediates yielded a smaller MSE only for a sufficiently large number of trajectories per replica.
Furthermore, in this system the optimal intermediates consistently dissipated more work than the linear ones as well as the logarithmic ones with $\alpha=1$.
For logarithmic intermediates with $\alpha=2$, the dissipated work can be larger or smaller than for optimal intermediates.

As a second system, we studied a three-state model of a double-well potential, in which the two energy minima exchange roles during the switching, mimicking barrier-crossing problems encountered in biophysics.
For this system, the optimal intermediates shared one feature: the transition-state energy varied non-monotonically, which likely eases the barrier crossing.
Moreover, optimal intermediates outperformed linear and logarithmic ones in all cases studied, with MSE reductions frequently exceeding an order of magnitude.
The improvement was largest for large differences in barrier heights and, in marked contrast to the two-state system, grew with the number of intermediates $N$.
Optimal intermediates also reduced the mean work compared with linear and logarithmic ones, most strongly in exactly those cases showing the largest MSE improvement: large differences in barrier heights and slow driving.

As a third system, we studied a potential well that is shifted to a new position in finite time, a toy model of optical-trap experiments.
For this system too, optimal intermediates always yielded a smaller MSE than linear and logarithmic ones, even for small values of $n$.
The reduction was largest for deep wells and slow driving; the mean work showed the same pattern.
By numerically solving the master equation, we showed that the optimal intermediates transport the probability distribution at almost constant speed, whereas linear and logarithmic intermediates lead to an accelerating motion.
In most cases, the optimal intermediates transport the probability distribution farther, so that it ends closer to
the final position of the well.

Comparing all systems studied here reveals several trends that may serve as heuristics for non-equilibrium free energy calculations of more realistic molecular systems.
First, the optimal intermediates always show distinct jumps in the energy levels at the initial and final times.
Generally, these jumps are more pronounced for shorter switching times, that is, for systems driven far from equilibrium.
Second, the MSE reduction achieved by optimal intermediates relative to linear and logarithmic interpolation is generally most pronounced for large changes in the energy landscape of a system (large $\Delta E$, $|\Delta _1-\Delta _2|$, or $\delta$). 
Third, for barrier-crossing problems, the barrier should be lowered rapidly and raised again later to accelerate barrier crossing.
Fourth, in contrast to previous suggestions \cite{schmiedl2007optimal, engel2023optimal}, minimizing dissipation does not guarantee faster convergence, and the dissipated work is not necessarily a good proxy for the MSE of non-equilibrium free energy estimates.

All results in this work were obtained for discrete model systems evolving under Glauber dynamics.
Whether the identified trends hold for other dynamics and more complex systems remains an open question; answering this question will require studying a broad range of more realistic systems.
However, our method for calculating optimal intermediate Hamiltonians relies on an exact solution of the tilted master equation and is therefore limited to systems with few discrete states.
In particular, our approach cannot be applied straightforwardly to systems arising in computational biophysics and soft matter, such as colloidal particles in continuous potential landscapes or many-body systems interacting through pair potentials.
Generalizing the optimization procedure introduced here to more realistic systems, for instance using differentiable simulations \cite{engel2023optimal}, is therefore a natural avenue for future work.
In any case, using optimal intermediates in free energy calculations for realistic systems will likely require iterative procedures \cite{schmiedl2007optimal, lindberg_optimizing, engel2023optimal}, because determining the exact optimal intermediates before the free energy calculation would be far more expensive than an accurate free energy calculation itself.
If such schemes prove successful, iteratively optimized intermediate Hamiltonians could become a key ingredient for improving the accuracy of non-equilibrium free energy calculations.

\begin{acknowledgments}
D.B. thanks Dr. Steffen Schultze for helpful discussions.
\end{acknowledgments}

\section*{Data Availability Statement}
The Python scripts and numerical data generated in this study will be made publicly available in a repository upon publication.

\section*{Conflict of Interest}
The authors have no conflicts to disclose.

\section*{Author Contributions}
D.B. developed the computational and theoretical framework, performed the numerical calculations, analyzed the results, and wrote the initial draft of the manuscript. 
H.G. conceived the study, acquired funding, contributed to the analysis, supervised the project, and reviewed and edited the manuscript. 
Both authors approved the final version of the manuscript.

\appendix

\section{Detailed Calculations for the Two-State Model with a Single Intermediate}
\label{sec:appendix_pert}
As stated in the main text, \autoref{eq:nonlinear_equation_toy_model} cannot be solved exactly. 
Here, we derive the asymptotic solutions of this equation for $E_{\mathrm{f}}\rightarrow 0$ and $E_{\mathrm{f}}\rightarrow \pm \infty$ using first-order perturbation theory \cite{bender1999advanced}.

\subsection{Limit of Small $E_{\mathrm{f}}$}
Let us first consider the case $\left|E_{\mathrm{f}}\right|\ll 1$.
With the small expansion parameter $\epsilon \equiv 2E_{\mathrm{f}}$, \autoref{eq:nonlinear_equation_toy_model} reads
\begin{align}
    G(\lambda^*) = G(-\lambda^*)e^{-\epsilon}.
    \label{eq:eq_perturbative}
\end{align}
To extract the asymptotic behavior for $\epsilon \rightarrow 0$, we expand the general solution $\lambda^*$ into a perturbation series:
\begin{align}
    \lambda^* = \lambda^*_0 + \epsilon \lambda_1^* + \mathcal{O}(\epsilon^2).
\end{align}
Inserting the expansion into \autoref{eq:eq_perturbative} and expanding up to linear order in $\epsilon$, we obtain
\begin{align}
\begin{split}
    &G(\lambda_0^*) + \epsilon \lambda_1^* G'(\lambda_0^*) + \mathcal{O}(\epsilon^2)\\ 
    =&\, G(-\lambda_0^*) - \epsilon\left[G(-\lambda_0^*)+\lambda_1^* G'(-\lambda_0^*)\right] + \mathcal{O}(\epsilon^2),
\end{split}
\end{align}
which leads to the equations
\begin{align}
    G(\lambda_0^*) &= G(-\lambda_0^*)
\end{align}
and
\begin{align}
    \lambda_1^* &= -\frac{G(-\lambda_0^*)}{G'(\lambda_0^*)+G'(-\lambda_0^*)}.
\end{align}
The first equation is solved by 
\begin{align}
    \lambda_0^* = 0
\end{align}
and the second equation yields
\begin{align}
    \lambda_1^* &= -\frac{G(0)}{2G'(0)} =  \frac{1}{4}.
\end{align}
Thus, the solution exhibits the following asymptotic behavior:
\begin{align}
    \lambda^* \sim \frac{E_{\mathrm{f}}}{2}, \quad \text{for }E_{\mathrm{f}} \to 0.
\end{align}

\subsection{Limit of Large $E_{\mathrm{f}}$}
In the limit $E_{\mathrm{f}}\gg 1$, $\epsilon \equiv e^{-2E_{\mathrm{f}}}\ll 1$ and \autoref{eq:nonlinear_equation_toy_model} becomes
\begin{align}
    G(\lambda^*) = \epsilon G(-\lambda^*).
    \label{eq:equation_large_E}
\end{align}
Expanding $\lambda^*$ into a perturbation series in $\epsilon$ and inserting this ansatz into \autoref{eq:equation_large_E}, we obtain
\begin{align}
        G(\lambda^*_0) + \epsilon G'(\lambda^*_0)\lambda^*_1 + \mathcal{O}(\epsilon^2) = \epsilon G(-\lambda_0^*) + \mathcal{O}(\epsilon^2),
\end{align}
which results in the equations
\begin{align}
        G(\lambda^*_0) &= 0
\end{align}
and
\begin{align}
        \lambda^*_1 &=  \frac{G(-\lambda_0^*)}{G'(\lambda^*_0)}.
        \label{eq:first_order_equation}
\end{align}
The zeroth-order equation is solved by
\begin{align}
        \lambda^*_0 = \sinh^{-1}(1).
\end{align}
Inserting this solution into \autoref{eq:first_order_equation} allows us to calculate $\lambda_1^*$:
\begin{align}
        \lambda^*_1 = \frac{G(-\sinh^{-1}(1))}{G'(\sinh^{-1}(1))} = -\left(3 \sqrt{2} + 4\right).
\end{align}
Combining $\lambda_0^*$ and $\lambda_1^*$, we obtain the asymptotic solution for $E_{\mathrm{f}}\to \infty$:
\begin{align}
    \lambda^* \sim \sinh^{-1}(1) -\left(3 \sqrt{2} + 4\right)\cdot e^{-2E_{\mathrm{f}}}, \quad \text{for } E_{\mathrm{f}} \to \infty.
\end{align}
An analogous perturbation expansion is possible for 
$E_{\mathrm{f}}\to -\infty$, where $\epsilon \equiv e^{2E_{\mathrm{f}}}\ll 1$ and \autoref{eq:nonlinear_equation_toy_model} becomes
\begin{align}
    G(-\lambda^*) = \epsilon G(\lambda^*).
\end{align}
Because this equation is equivalent to \autoref{eq:equation_large_E} under the substitution $\lambda^*\to -\lambda^*$, the asymptotic solution reads
\begin{align}
    \lambda^* \sim -\sinh^{-1}(1) +\left(3 \sqrt{2} + 4\right)\cdot e^{2E_{\mathrm{f}}}, \quad \text{for } E_{\mathrm{f}} \to -\infty.
\end{align}

\bibliography{bibliography}

\makeatletter\@input{xx.tex}\makeatother

\end{document}

\title{Supplemental Material: Optimal Intermediate Hamiltonians for Non-Equilibrium Free Energy Calculations: A Numerical Study of Markov Models}

\author{David Beyer}
\email{david.beyer@mpinat.mpg.de}
\affiliation{Department of Theoretical and
Computational Biophysics, Max Planck Institute for Multidisciplinary Sciences, D-37077 Göttingen, Germany}

\author{Helmut Grubm\"{u}ller}
\email{hgrubmu@gwdg.de}
\affiliation{Department of Theoretical and
Computational Biophysics, Max Planck Institute for Multidisciplinary Sciences, D-37077 Göttingen, Germany}
\maketitle

\section{Additional Figures}

We provide the following additional figures.

\subsection{Two-State Model}
\begin{figure*}[ht]
\centering
\includegraphics[width=\textwidth]{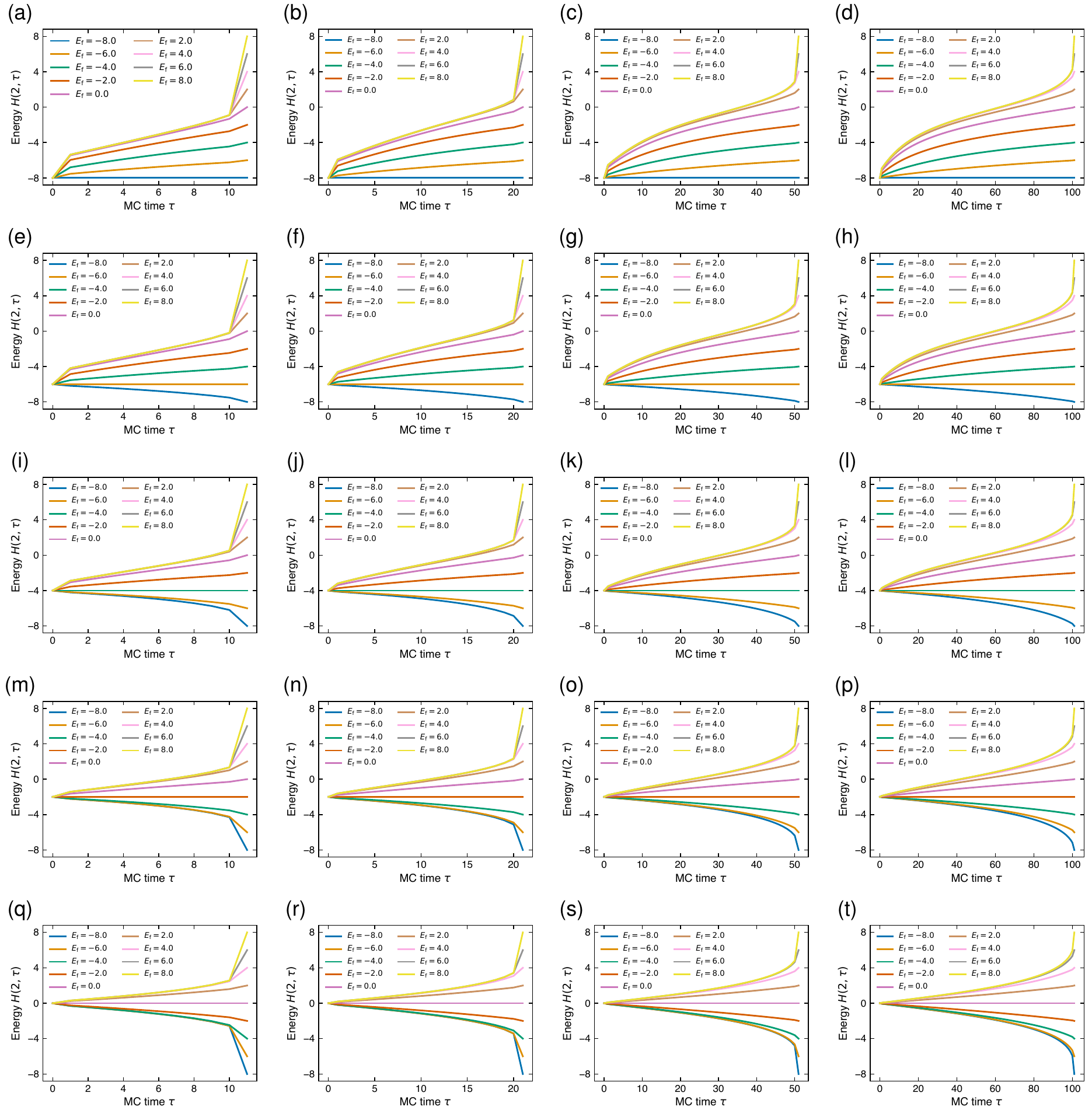}
\caption{\label{fig:exhaustive_protocols_two_state_1} 
Optimal intermediate Hamiltonians $H(2,\tau)$ of the two-state system for different initial energies $E_{\mathrm{i}}$ and switching durations $N$.
(a)--(d): $E_{\mathrm{i}}=-8$, (e)--(h): $E_{\mathrm{i}}=-6$ (i)--(l): $E_{\mathrm{i}}=-4$, (m)--(p): $E_{\mathrm{i}}=-2$, (q)--(t): $E_{\mathrm{i}}=0$.}
\end{figure*}

\begin{figure*}[ht]
\centering
\includegraphics[width=\textwidth]{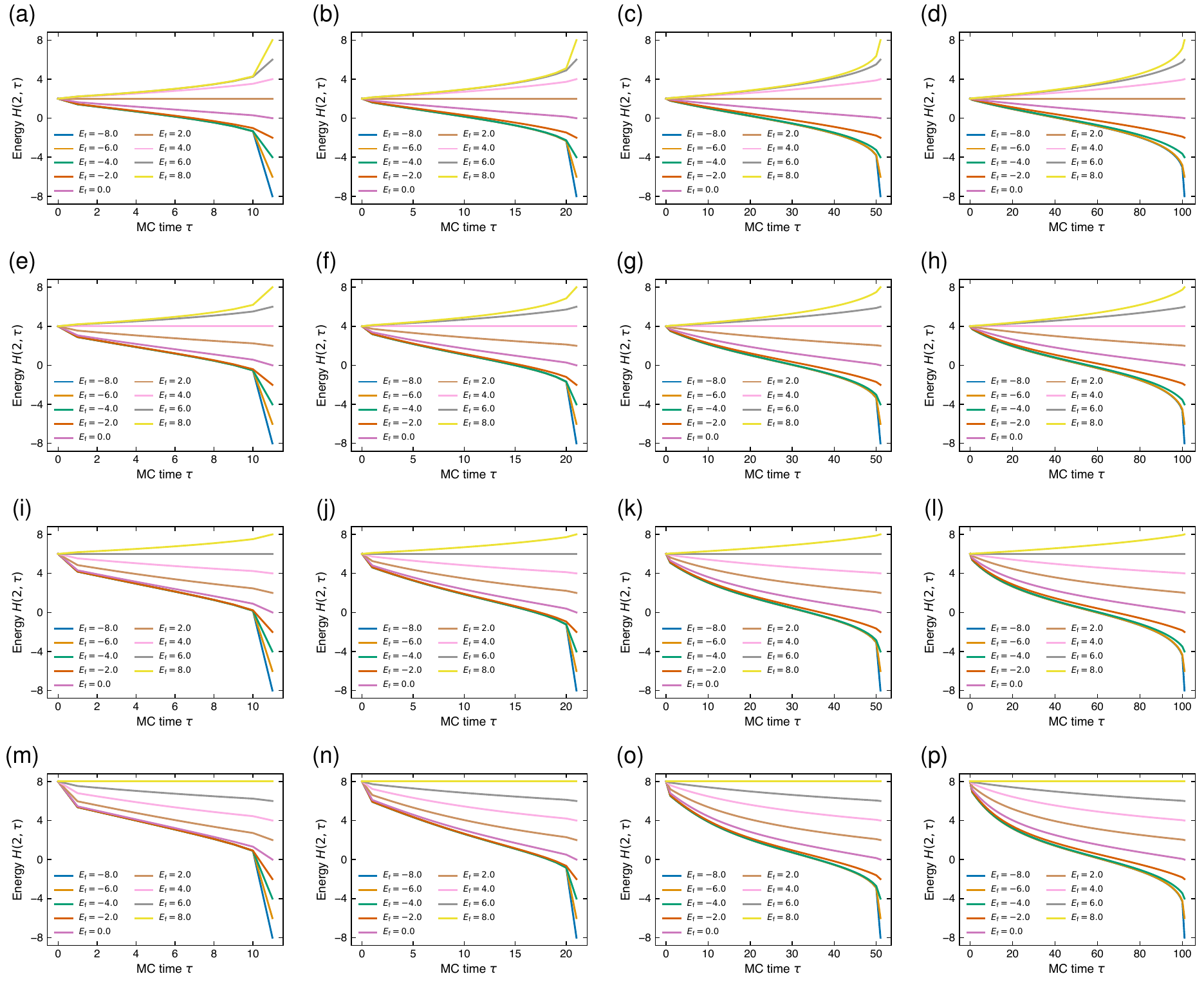}
\caption{\label{fig:exhaustive_protocols_two_state_2} 
Optimal intermediate Hamiltonians $H(2,\tau)$ of the two-state system for different initial energies $E_{\mathrm{i}}$ and switching durations $N$.
(a)--(d): $E_{\mathrm{i}}=2$, (e)--(h): $E_{\mathrm{i}}=4$ (i)--(l): $E_{\mathrm{i}}=6$, (m)--(p): $E_{\mathrm{i}}=8$.}
\end{figure*}

\begin{figure*}[ht]
\centering
\includegraphics[width=\textwidth]{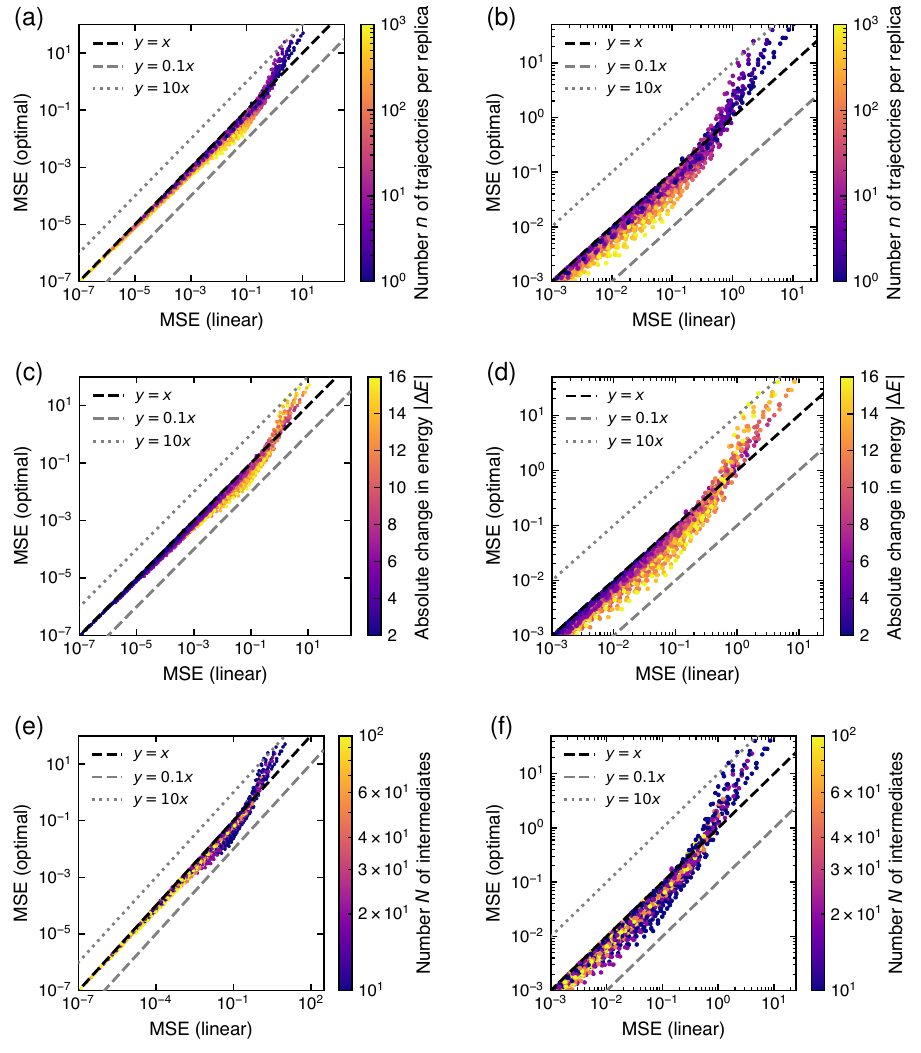}
\caption{\label{fig:correlation_plots_mse_two_state_esi} 
Correlation plots of the MSE for linear and optimal intermediates of the two-state system.
Error bars corresponding to the standard error of the mean are included, but smaller than the symbol size.
(a) and (b): Correlation plots colored according to the number $n$ of trajectories per replica.
(c) and (d): Correlation plots colored according to the absolute change in energy $\Delta E$.
(e) and (f): Correlation plots colored according to the number $N$ of intermediates.}
\end{figure*}

\begin{figure*}[ht]
\centering
\includegraphics[width=\textwidth]{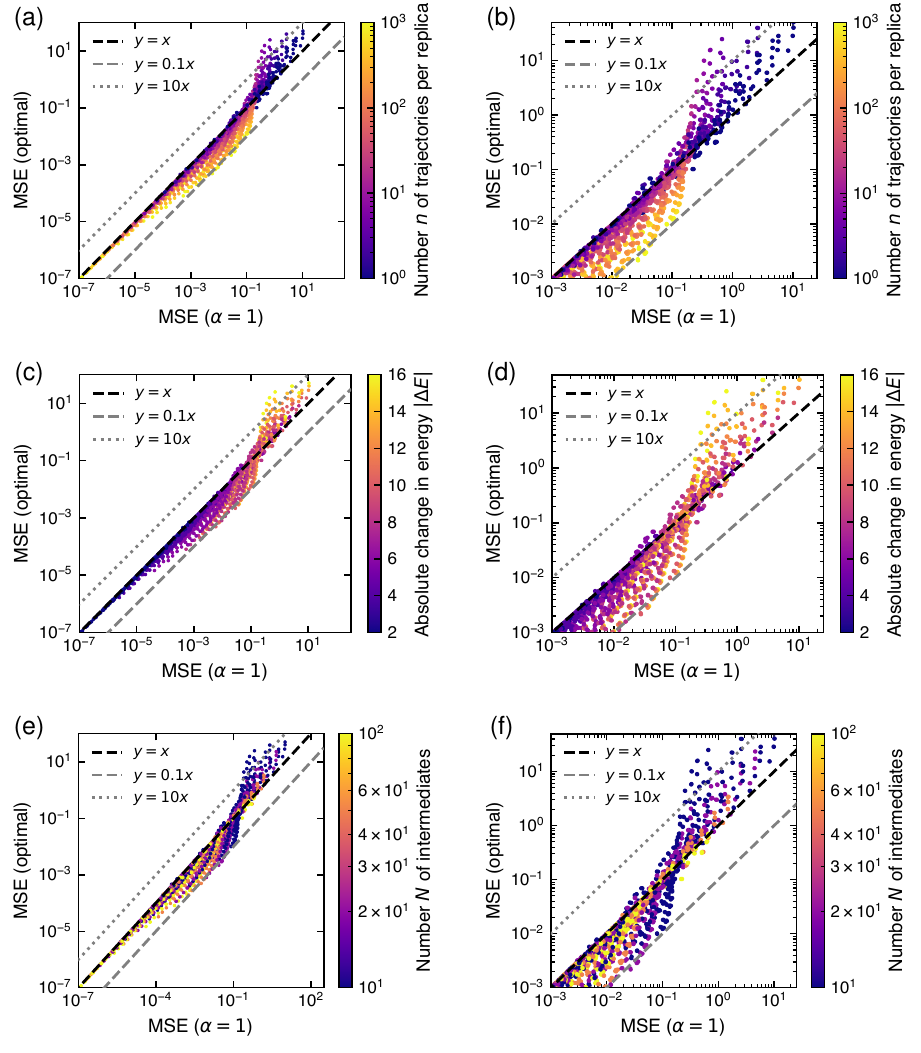}
\caption{\label{fig:correlation_plots_mse_two_state_log_linear_esi} 
Correlation plots of the MSE for logarithmic ($\alpha=1$) and optimal intermediates of the two-state system.
Error bars corresponding to the standard error of the mean are included, but smaller than the symbol size.
(a) and (b): Correlation plots colored according to the number $n$ of trajectories per replica.
(c) and (d): Correlation plots colored according to the absolute change in energy $\Delta E$.
(e) and (f): Correlation plots colored according to the number $N$ of intermediates.}
\end{figure*}

\begin{figure*}[ht]
\centering
\includegraphics[width=\textwidth]{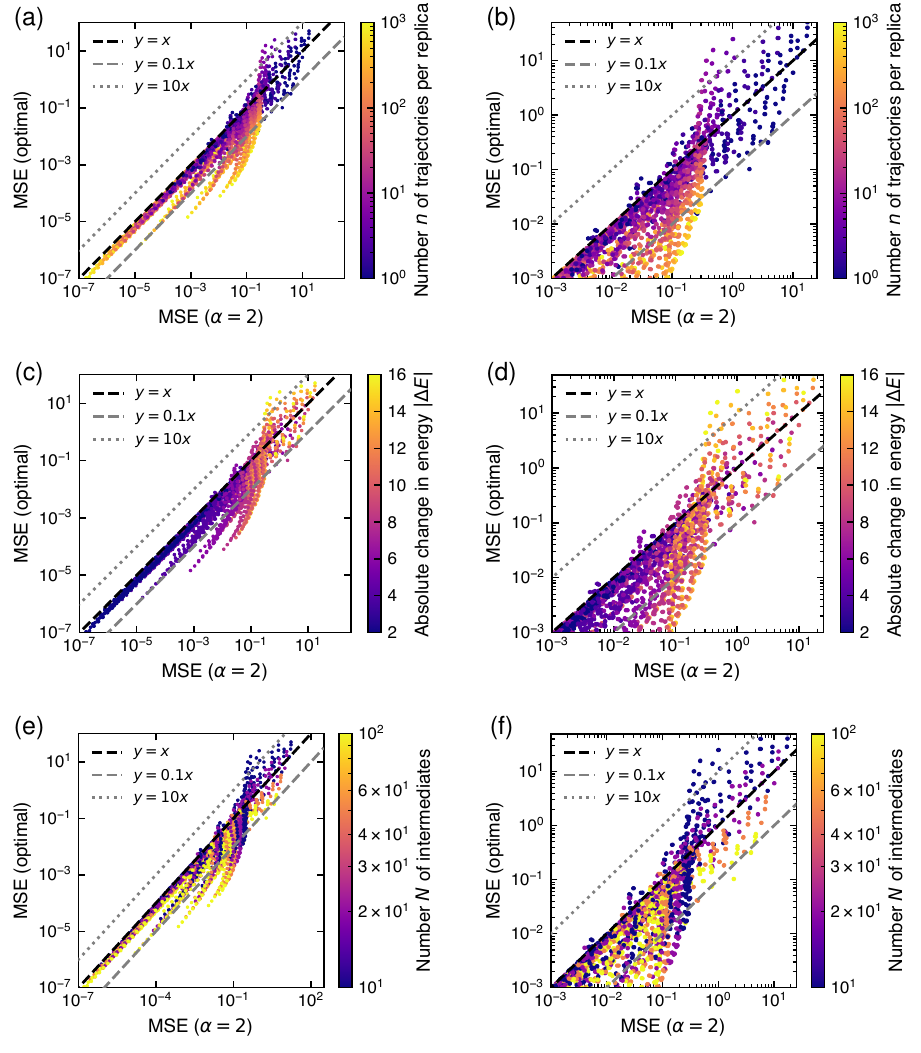}
\caption{\label{fig:correlation_plots_mse_two_state_avi_esi} 
Correlation plots of the MSE for logarithmic ($\alpha=2$) and optimal intermediates of the two-state system.
Error bars corresponding to the standard error of the mean are included, but smaller than the symbol size.
(a) and (b): Correlation plots colored according to the number $n$ of trajectories per replica.
(c) and (d): Correlation plots colored according to the absolute change in energy $\Delta E$.
(e) and (f): Correlation plots colored according to the number $N$ of intermediates.}
\end{figure*}

\begin{figure*}[ht]
\centering
\includegraphics[width=0.8\textwidth]{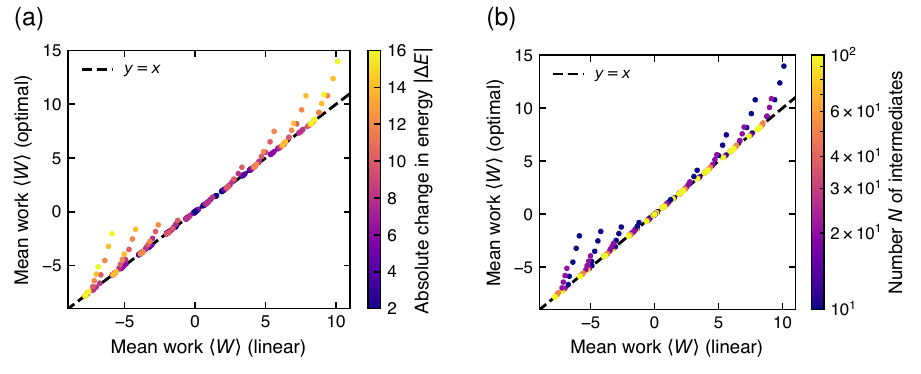}
\caption{\label{fig:correlation_plots_work_two_state_esi} 
Correlation plots of the mean work for linear and optimal intermediates of the two-state system.
Error bars corresponding to the standard error of the mean are included, but smaller than the symbol size.
(a): Correlation plot colored according to the absolute change in energy $\Delta E$.
(b): Correlation plot colored according to the number $N$ of intermediates.}
\end{figure*}

\begin{figure*}[ht]
\centering
\includegraphics[width=0.8\textwidth]{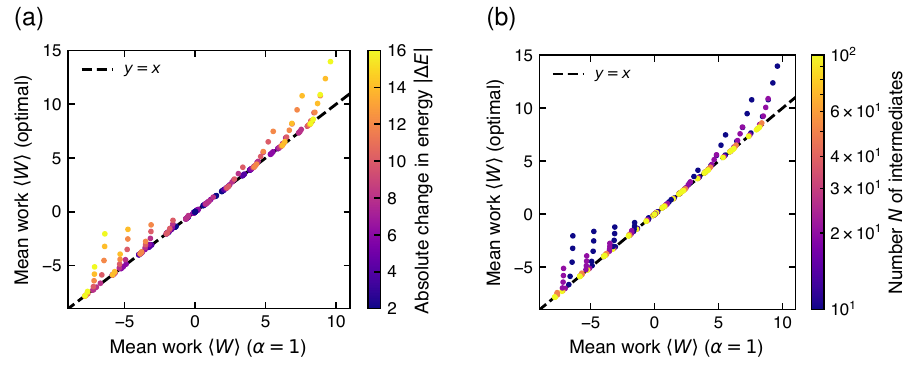}
\caption{\label{fig:correlation_plots_work_two_state_log_linear_esi} 
Correlation plots of the mean work for logarithmic ($\alpha=1$) and optimal intermediates of the two-state system.
Error bars corresponding to the standard error of the mean are included, but smaller than the symbol size.
(a): Correlation plot colored according to the absolute change in energy $\Delta E$.
(b): Correlation plot colored according to the number $N$ of intermediates.}
\end{figure*}

\begin{figure*}[ht]
\centering
\includegraphics[width=0.8\textwidth]{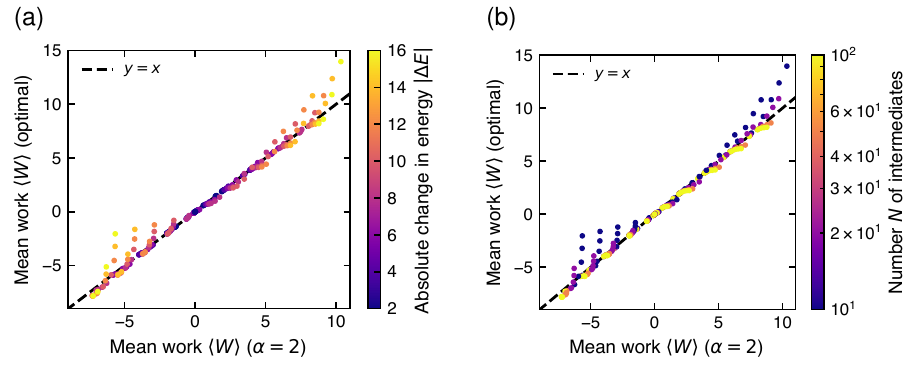}
\caption{\label{fig:correlation_plots_work_two_state_avi_esi} 
Correlation plots of the mean work for logarithmic ($\alpha=1$) and optimal intermediates of the two-state system.
Error bars corresponding to the standard error of the mean are included, but smaller than the symbol size.
(a): Correlation plot colored according to the absolute change in energy $\Delta E$.
(b): Correlation plot colored according to the number $N$ of intermediates.}
\end{figure*}

\begin{itemize}
    \item Plots of the optimal intermediate Hamiltonians for the two-state system (\autoref{fig:exhaustive_protocols_two_state_1} and \autoref{fig:exhaustive_protocols_two_state_2})
    \item Correlation plots of the MSE for linear and optimal intermediates of the two-state system (\autoref{fig:correlation_plots_mse_two_state_esi})
    \item Correlation plots of the MSE for logarithmic ($\alpha=1$) and optimal intermediates of the two-state system (\autoref{fig:correlation_plots_mse_two_state_log_linear_esi})
    \item Correlation plots of the MSE for logarithmic ($\alpha=2$) and optimal intermediates of the two-state system (\autoref{fig:correlation_plots_mse_two_state_avi_esi})
    \item Correlation plots of the mean work for linear and optimal intermediates of the two-state system 
    (\autoref{fig:correlation_plots_work_two_state_esi})
    \item Correlation plots of the mean work for logarithmic ($\alpha=1$) and optimal intermediates of the two-state system 
    (\autoref{fig:correlation_plots_work_two_state_log_linear_esi})
    \item Correlation plots of the mean work for logarithmic ($\alpha=2$) and optimal intermediates of the two-state system 
    (\autoref{fig:correlation_plots_work_two_state_avi_esi})
\end{itemize}

\subsection{Double-Well Potential}
\begin{figure*}[ht]
\centering
\includegraphics[width=\textwidth]{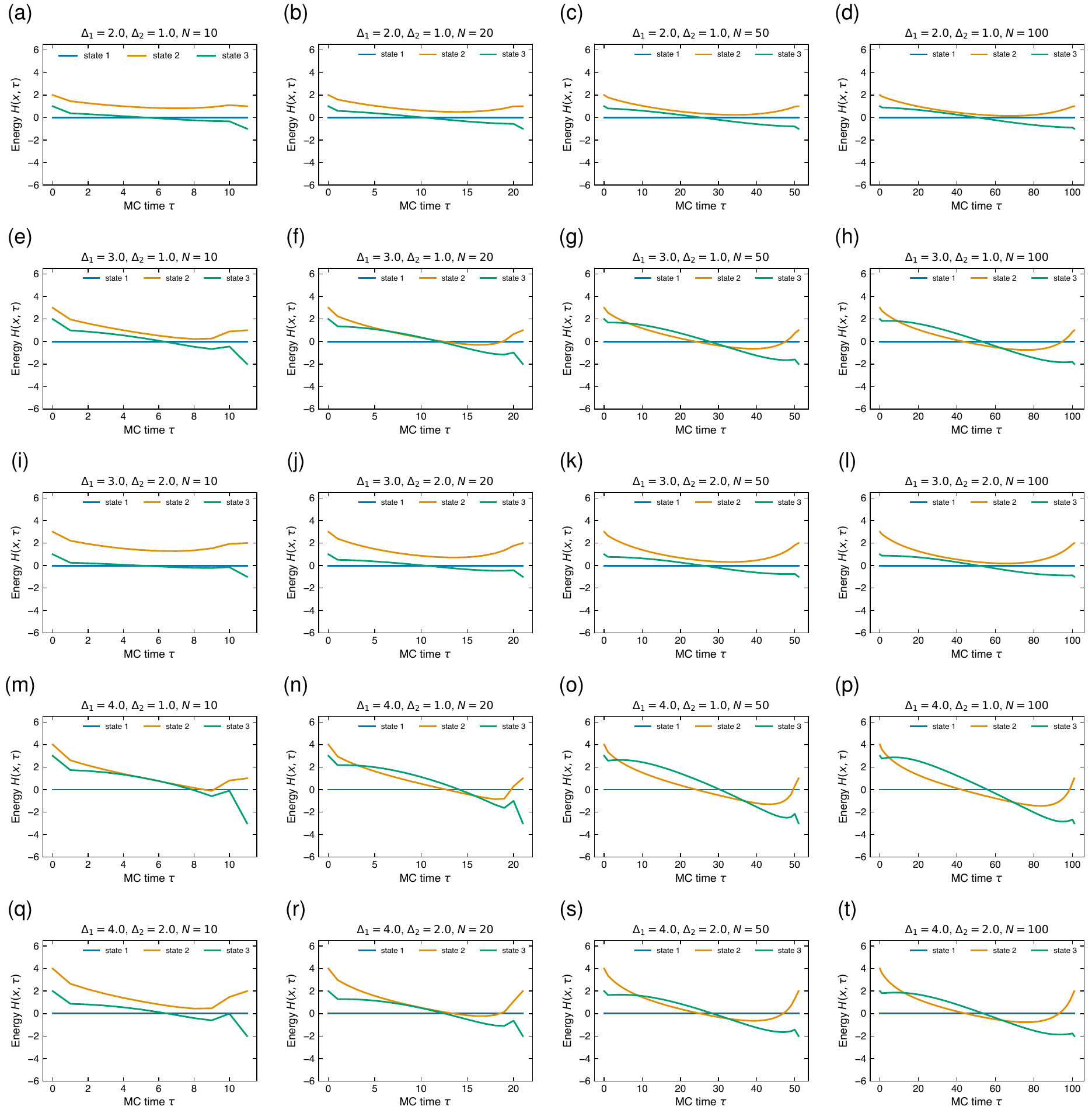}
\caption{\label{fig:exhaustive_protocols_three_state_1} 
Optimal intermediate Hamiltonians $H(x,\tau)$ of the double-well potential with different barrier heights $\Delta_1$, $\Delta_2$ and numbers of intermediates $N$.}
\end{figure*}

\begin{figure*}[ht]
\centering
\includegraphics[width=\textwidth]{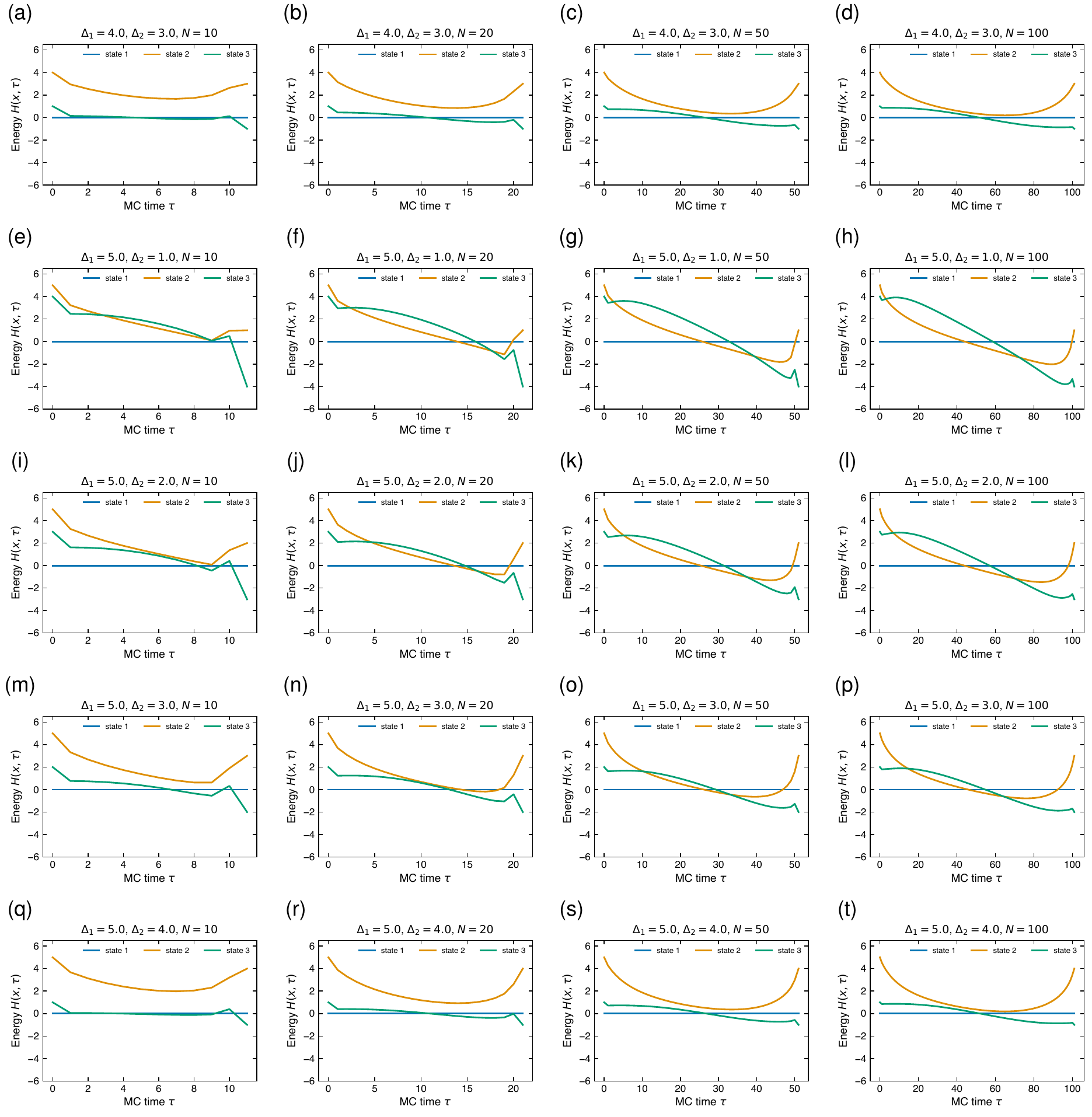}
\caption{\label{fig:exhaustive_protocols_three_state_2} 
Optimal intermediate Hamiltonians $H(x,\tau)$ of the double-well potential with different barrier heights $\Delta_1$, $\Delta_2$ and numbers of intermediates $N$.}
\end{figure*}

\begin{figure*}[ht]
\centering
\includegraphics[width=\textwidth]{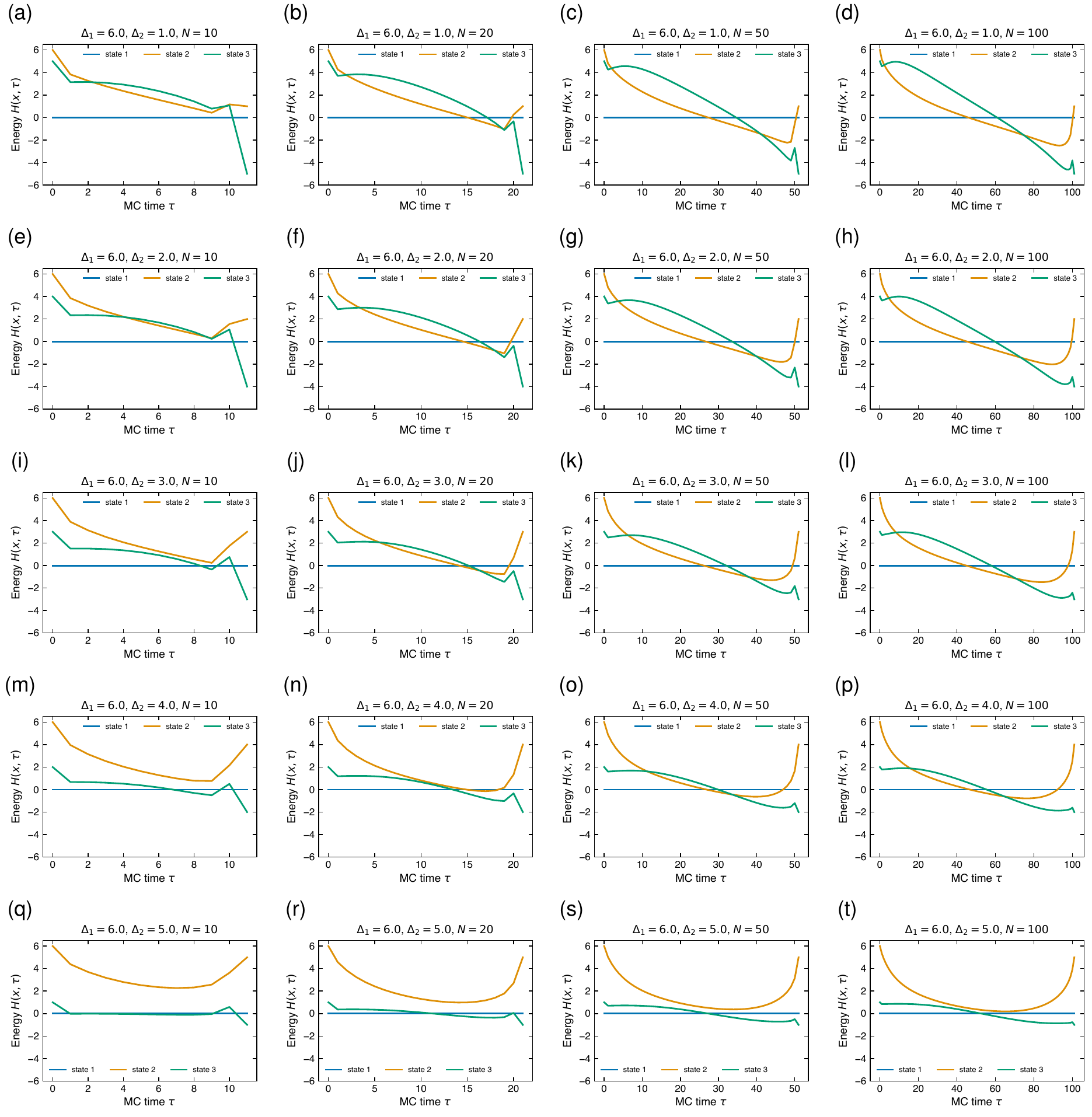}
\caption{\label{fig:exhaustive_protocols_three_state_3} 
Optimal intermediate Hamiltonians $H(x,\tau)$ of the double-well potential with different barrier heights $\Delta_1$, $\Delta_2$ and numbers of intermediates $N$.}
\end{figure*}

\begin{figure*}[ht]
\centering
\includegraphics[width=0.8\textwidth]{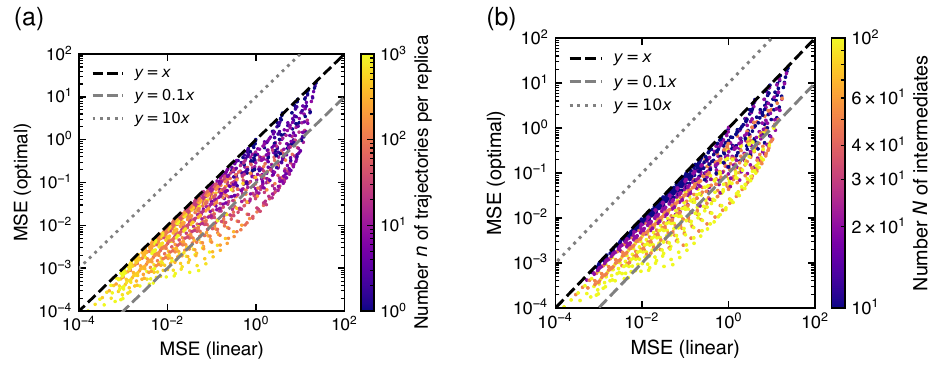}
\caption{\label{fig:correlation_plots_mse_three_state_esi} 
Correlation plots of the MSE for linear and optimal intermediates of the double-well potential. 
Error bars corresponding to the standard error of the mean are included, but smaller than the symbol size.
(a): Correlation plot colored according to the number $n$ of trajectories per replica.
(b): Correlation plot colored according to the number $N$ of intermediates.}
\end{figure*}

\begin{figure*}[ht]
\centering
\includegraphics[width=\textwidth]{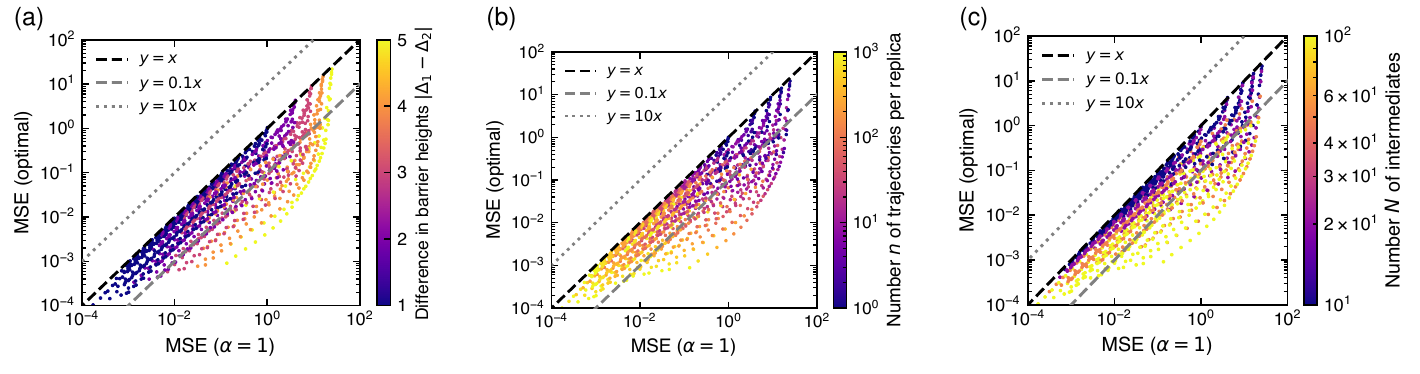}
\caption{\label{fig:correlation_plots_mse_three_state_log_linear_esi} 
Correlation plots of the MSE for logarithmic ($\alpha=1$) and optimal intermediates of the double-well potential. 
Error bars corresponding to the standard error of the mean are included, but smaller than the symbol size.
(a): Correlation plot colored according to the difference in barrier heights $|\Delta_1-\Delta_2|$.
(b): Correlation plot colored according to the number $n$ of trajectories per replica.
(c): Correlation plot colored according to the number $N$ of intermediates.}
\end{figure*}

\begin{figure*}[ht]
\centering
\includegraphics[width=\textwidth]{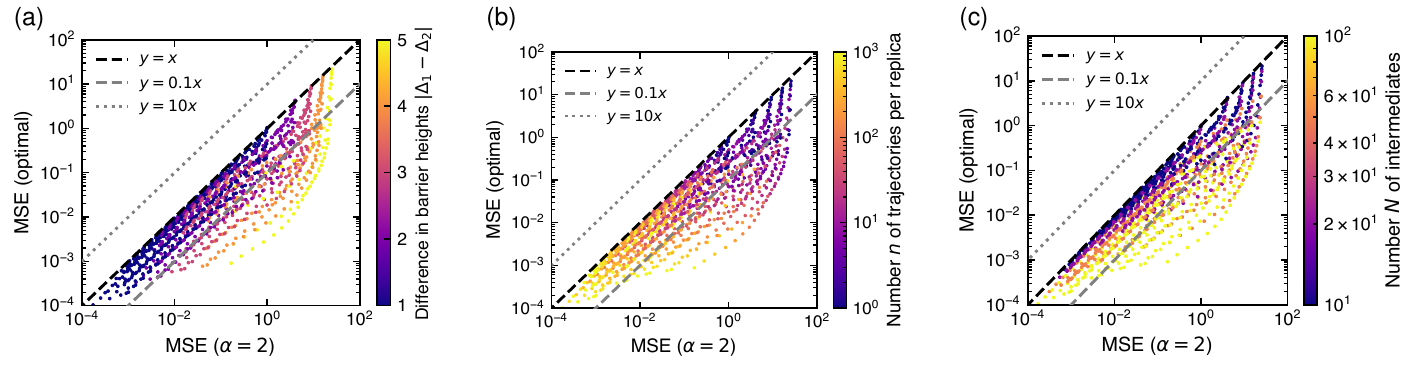}
\caption{\label{fig:correlation_plots_mse_three_state_avi_esi} 
Correlation plots of the MSE for logarithmic ($\alpha=2$) and optimal intermediates of the double-well potential. 
Error bars corresponding to the standard error of the mean are included, but smaller than the symbol size.
Note that these plots closely resemble those in \autoref{fig:correlation_plots_mse_three_state_log_linear_esi}, yet are not identical.
(a): Correlation plot colored according to the difference in barrier heights $|\Delta_1-\Delta_2|$.
(b): Correlation plot colored according to the number $n$ of trajectories per replica.
(c): Correlation plot colored according to the number $N$ of intermediates.}
\end{figure*}

\begin{figure*}[ht]
\centering
\includegraphics[width=0.8\textwidth]{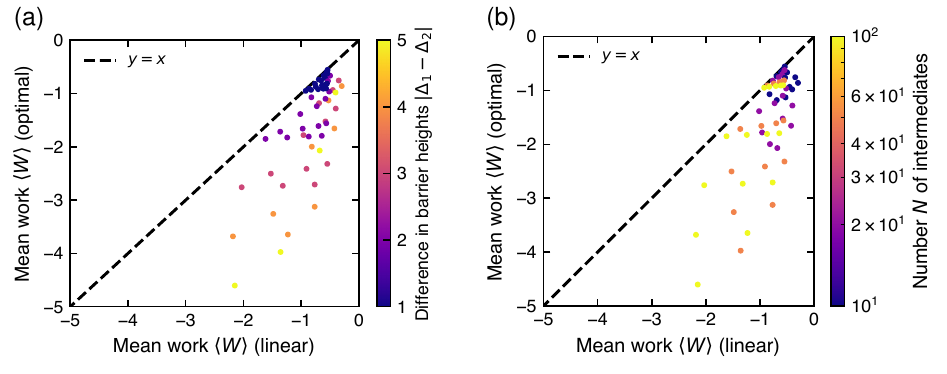}
\caption{\label{fig:correlation_plots_work_three_state_esi} 
Correlation plots of the mean work for linear and optimal intermediates of the double-well potential.
Error bars corresponding to the standard error of the mean are included, but smaller than the symbol size.
(a): Correlation plot colored according to the barrier height $\Delta_1$.
(b): Correlation plot colored according to the number $n$ of trajectories per replica.}
\end{figure*}

\begin{figure*}[ht]
\centering
\includegraphics[width=0.8\textwidth]{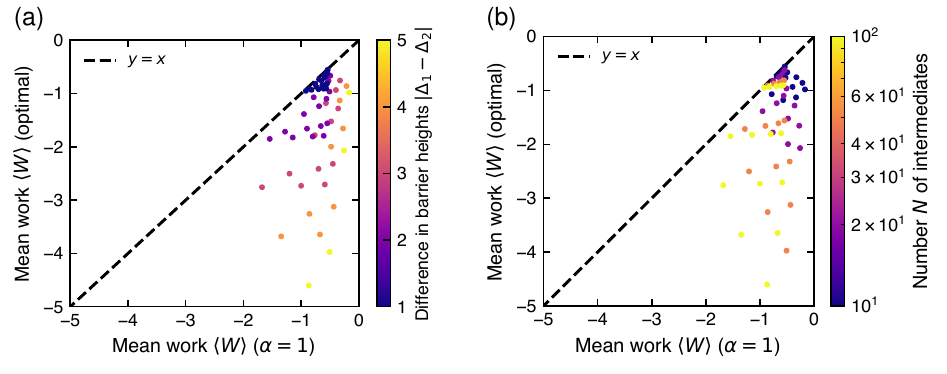}
\caption{\label{fig:correlation_plots_work_three_state_log_linear_esi} 
Correlation plots of the mean work for logarithmic ($\alpha=1$) and optimal intermediates of the double-well potential.
Error bars corresponding to the standard error of the mean are included, but smaller than the symbol size.
(a): Correlation plot colored according to the barrier height $\Delta_1$.
(b): Correlation plot colored according to the number $n$ of trajectories per replica.}
\end{figure*}

\begin{figure*}[ht]
\centering
\includegraphics[width=0.8\textwidth]{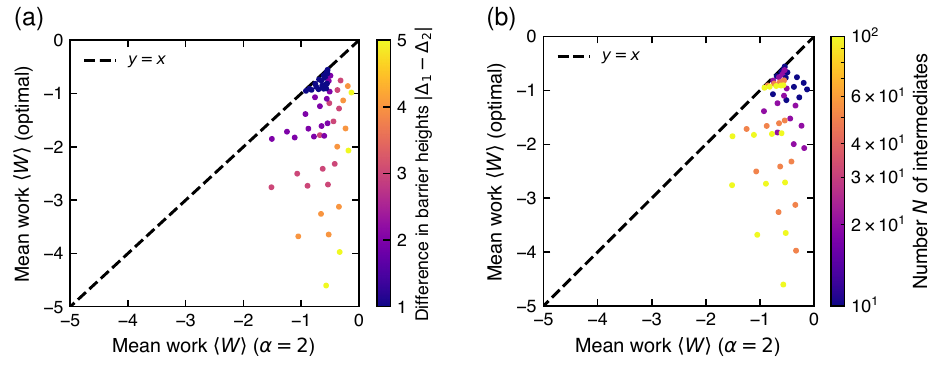}
\caption{\label{fig:correlation_plots_work_three_state_avi_esi} 
Correlation plots of the mean work for logarithmic ($\alpha=2$) and optimal intermediates of the double-well potential.
Error bars corresponding to the standard error of the mean are included, but smaller than the symbol size.
Note that these plots closely resemble those in \autoref{fig:correlation_plots_work_three_state_log_linear_esi}, yet are not identical.
(a): Correlation plot colored according to the barrier height $\Delta_1$.
(b): Correlation plot colored according to the number $n$ of trajectories per replica.}
\end{figure*}

\begin{itemize}
    \item Plots of the optimal intermediate Hamiltonians for the double-well potential (\autoref{fig:exhaustive_protocols_three_state_1}--\autoref{fig:exhaustive_protocols_three_state_3})
    \item Correlation plots of the MSE for linear and optimal intermediates of the double-well potential (\autoref{fig:correlation_plots_mse_three_state_esi})
    \item Correlation plots of the MSE for logarithmic ($\alpha=1$) and optimal intermediates of the double-well potential (\autoref{fig:correlation_plots_mse_three_state_log_linear_esi})
    \item Correlation plots of the MSE for logarithmic ($\alpha=2$) and optimal intermediates of the double-well potential (\autoref{fig:correlation_plots_mse_three_state_avi_esi})
    \item Correlation plots of the mean work for linear and optimal intermediates of the double-well potential (\autoref{fig:correlation_plots_work_three_state_esi})
    \item Correlation plots of the mean work for logarithmic ($\alpha=1$) and optimal intermediates of the double-well potential (\autoref{fig:correlation_plots_work_three_state_log_linear_esi})
    \item Correlation plots of the mean work for logarithmic ($\alpha=2$) and optimal intermediates of the double-well potential (\autoref{fig:correlation_plots_work_three_state_avi_esi})
\end{itemize}

\subsection{Shifted Potential Well}
\begin{figure*}[ht]
\centering
\includegraphics[width=\textwidth]{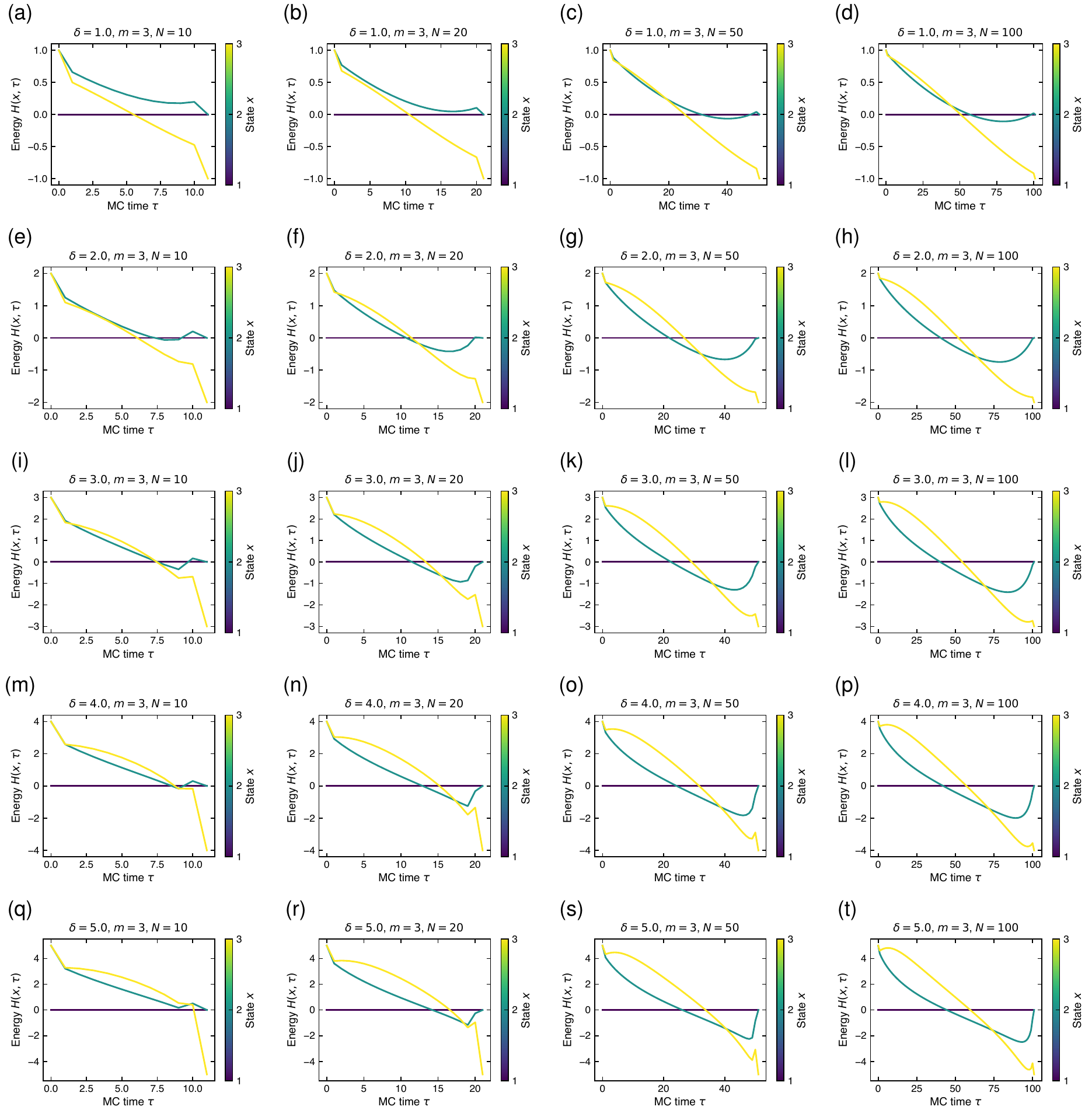}
\caption{\label{fig:exhaustive_protocols_pulling_single_well_1} 
Optimal intermediate Hamiltonians $H(x,\tau)$ of the shifted potential well with $m=3$ states and different depths $\delta$ and numbers of intermediates $N$.}
\end{figure*}

\begin{figure*}[ht]
\centering
\includegraphics[width=\textwidth]{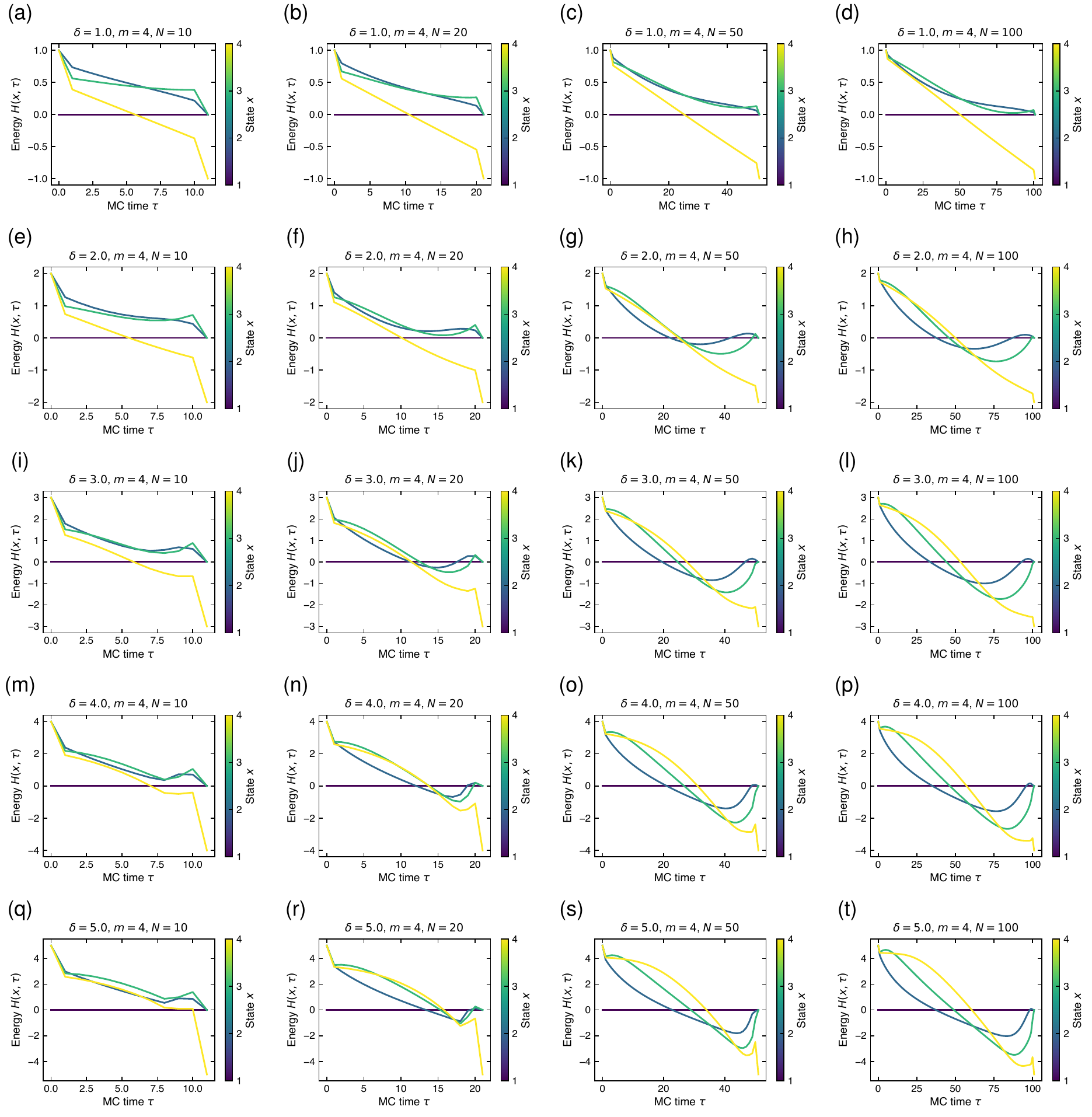}
\caption{\label{fig:exhaustive_protocols_pulling_single_well_2} 
Optimal intermediate Hamiltonians $H(x,\tau)$ of the shifted potential well with $m=4$ states and different depths $\delta$ and numbers of intermediates $N$.}
\end{figure*}

\begin{figure*}[ht]
\centering
\includegraphics[width=\textwidth]{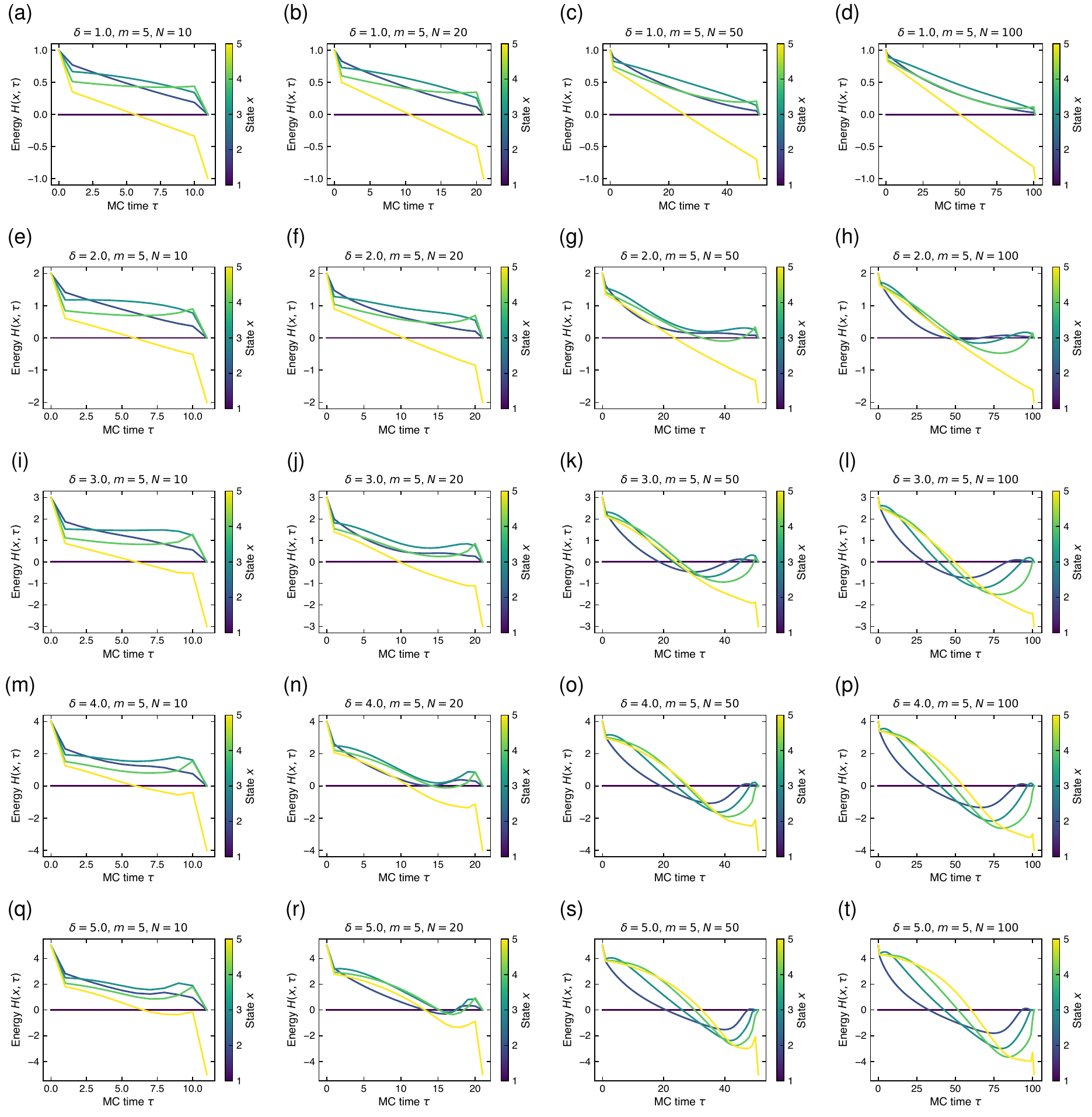}
\caption{\label{fig:exhaustive_protocols_pulling_single_well_3} 
Optimal intermediate Hamiltonians $H(x,\tau)$ of the shifted potential well with $m=5$ states and different depths $\delta$ and numbers of intermediates $N$.}
\end{figure*}

\begin{figure*}[ht]
\centering
\includegraphics[width=\textwidth]{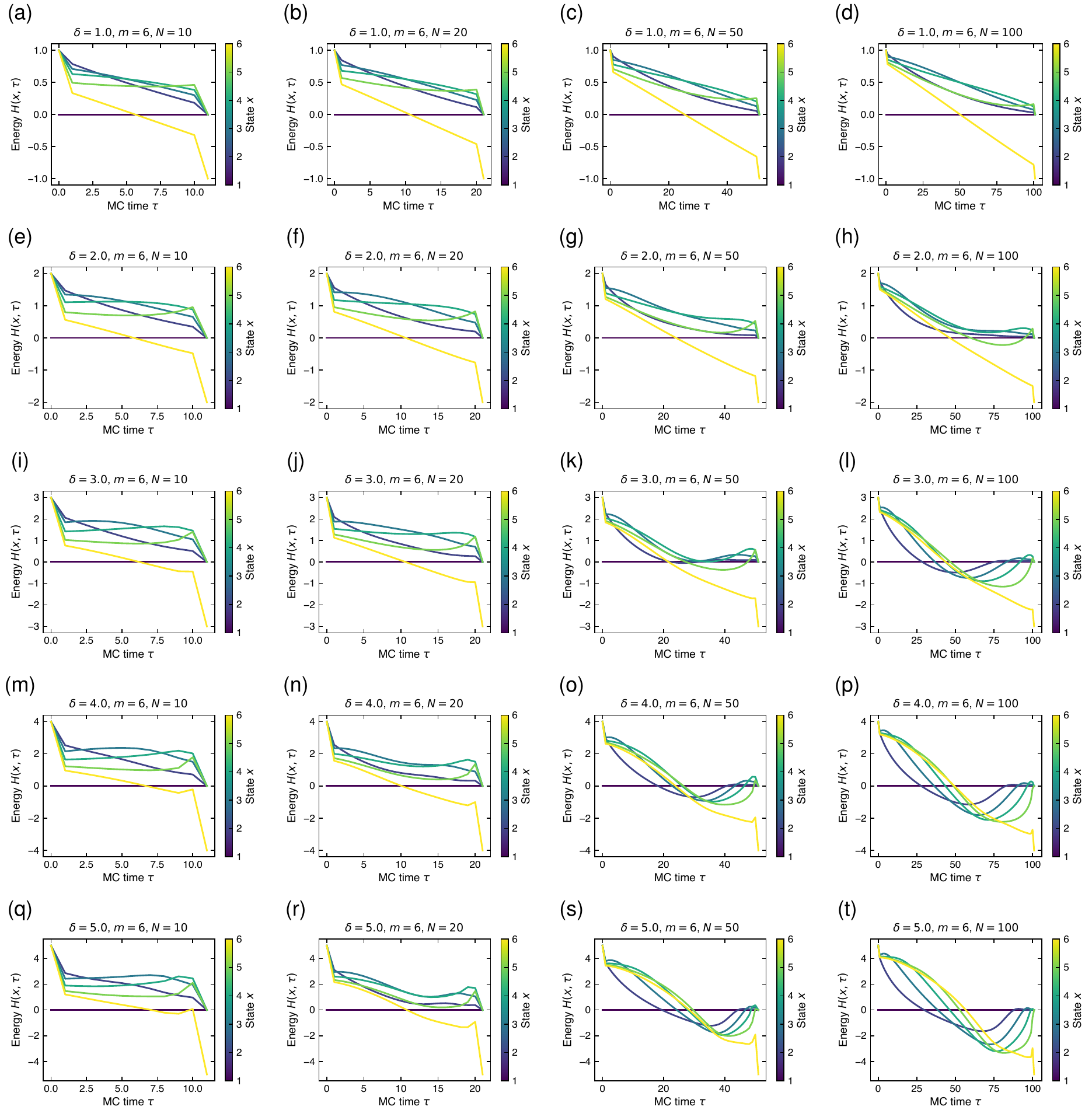}
\caption{\label{fig:exhaustive_protocols_pulling_single_well_4} 
Optimal intermediate Hamiltonians $H(x,\tau)$ of the shifted potential well with $m=6$ states and different depths $\delta$ and numbers of intermediates $N$.}
\end{figure*}

\begin{figure*}[ht]
\centering
\includegraphics[width=\textwidth]{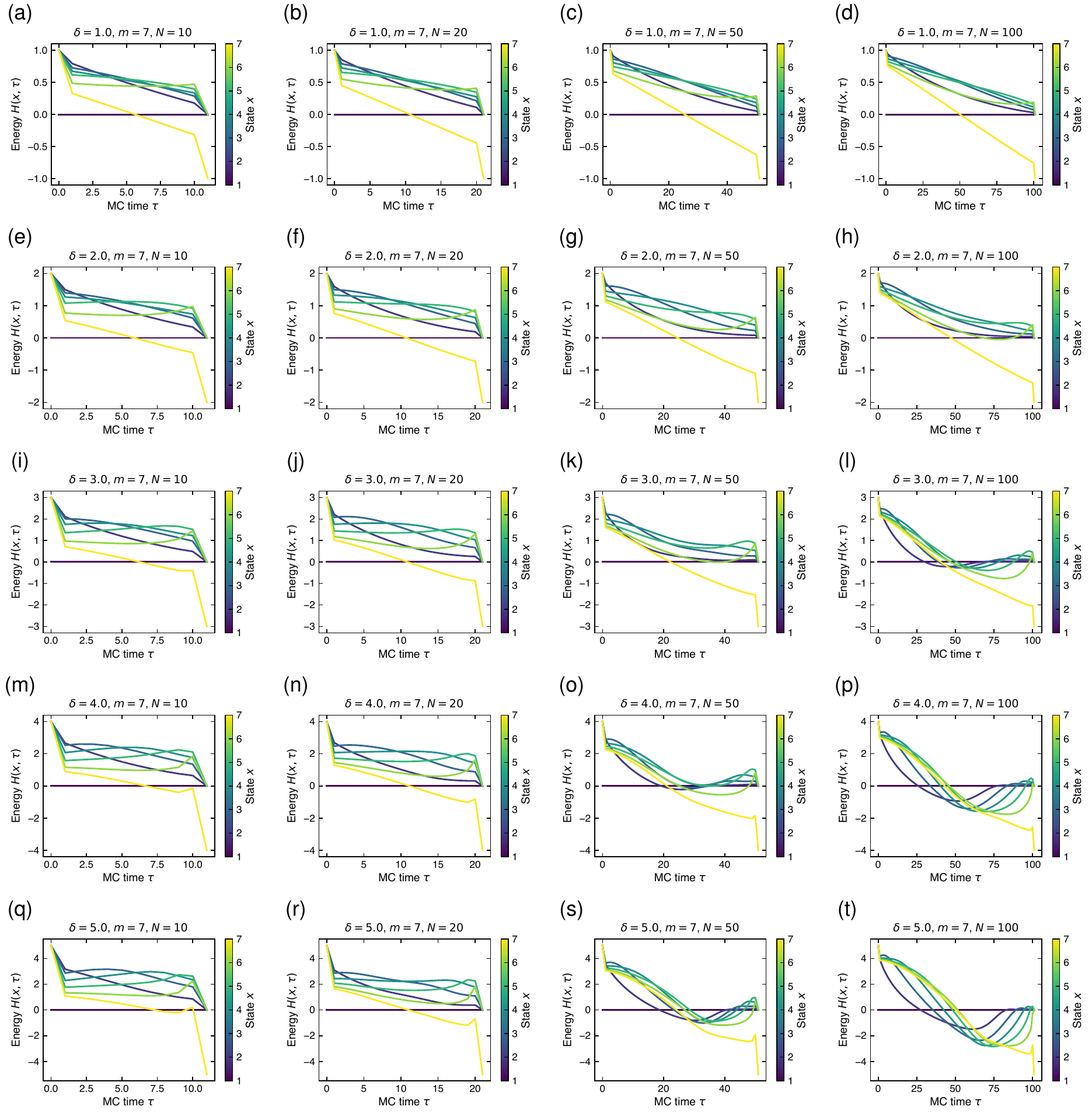}
\caption{\label{fig:exhaustive_protocols_pulling_single_well_5} 
Optimal intermediate Hamiltonians $H(x,\tau)$ of the shifted potential well with $m=7$ states and different depths $\delta$ and numbers of intermediates $N$.}
\end{figure*}

\begin{figure*}[ht]
\centering
\includegraphics[width=\textwidth]{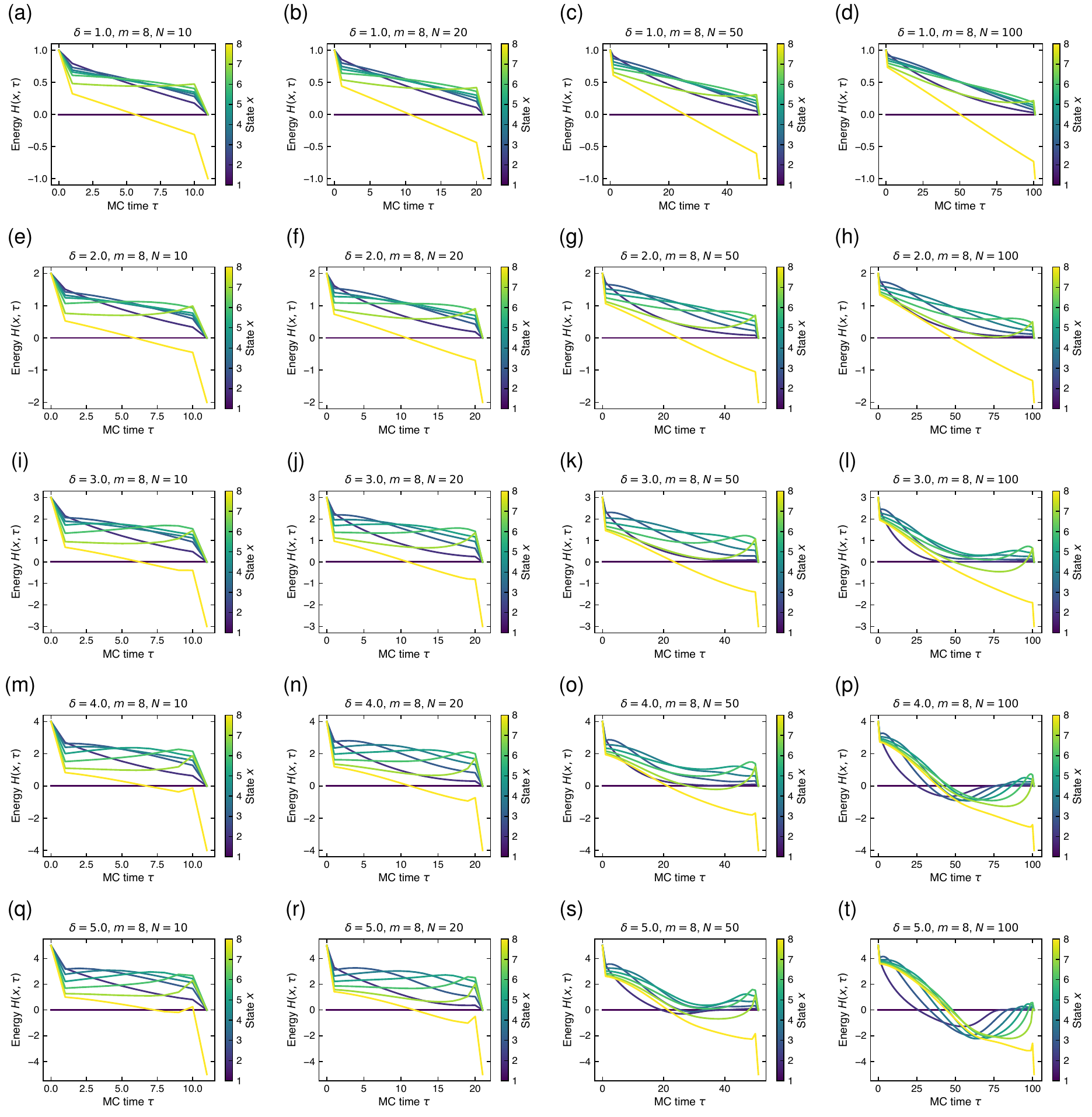}
\caption{\label{fig:exhaustive_protocols_pulling_single_well_6} 
Optimal intermediate Hamiltonians $H(x,\tau)$ of the shifted potential well with $m=8$ states and different depths $\delta$ and numbers of intermediates $N$.}
\end{figure*}

\begin{figure*}[ht]
\centering
\includegraphics[width=\textwidth]{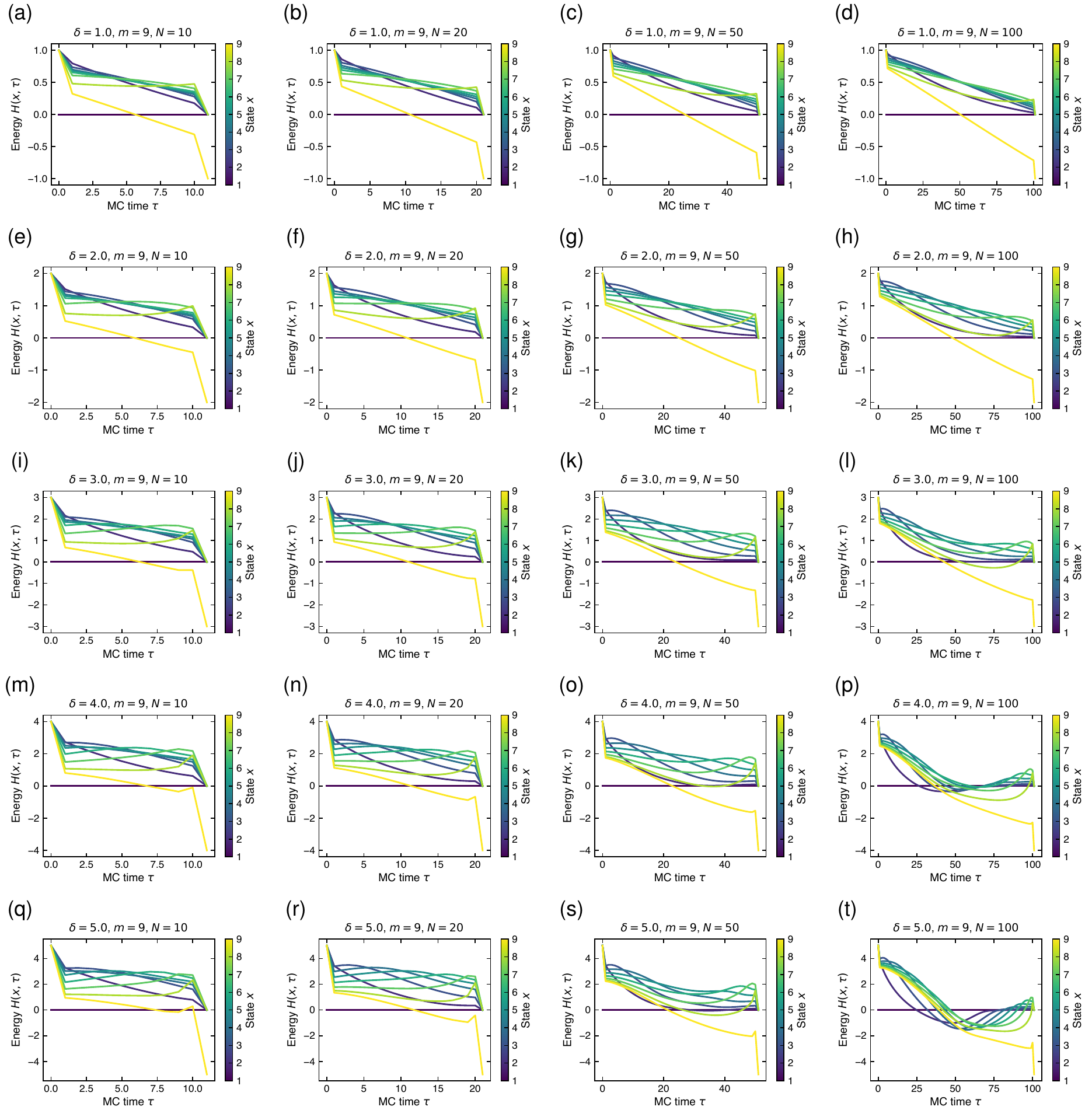}
\caption{\label{fig:exhaustive_protocols_pulling_single_well_7} 
Optimal intermediate Hamiltonians $H(x,\tau)$ of the shifted potential well with $m=9$ states and different depths $\delta$ and numbers of intermediates $N$.}
\end{figure*}

\begin{figure*}[ht]
\centering
\includegraphics[width=\textwidth]{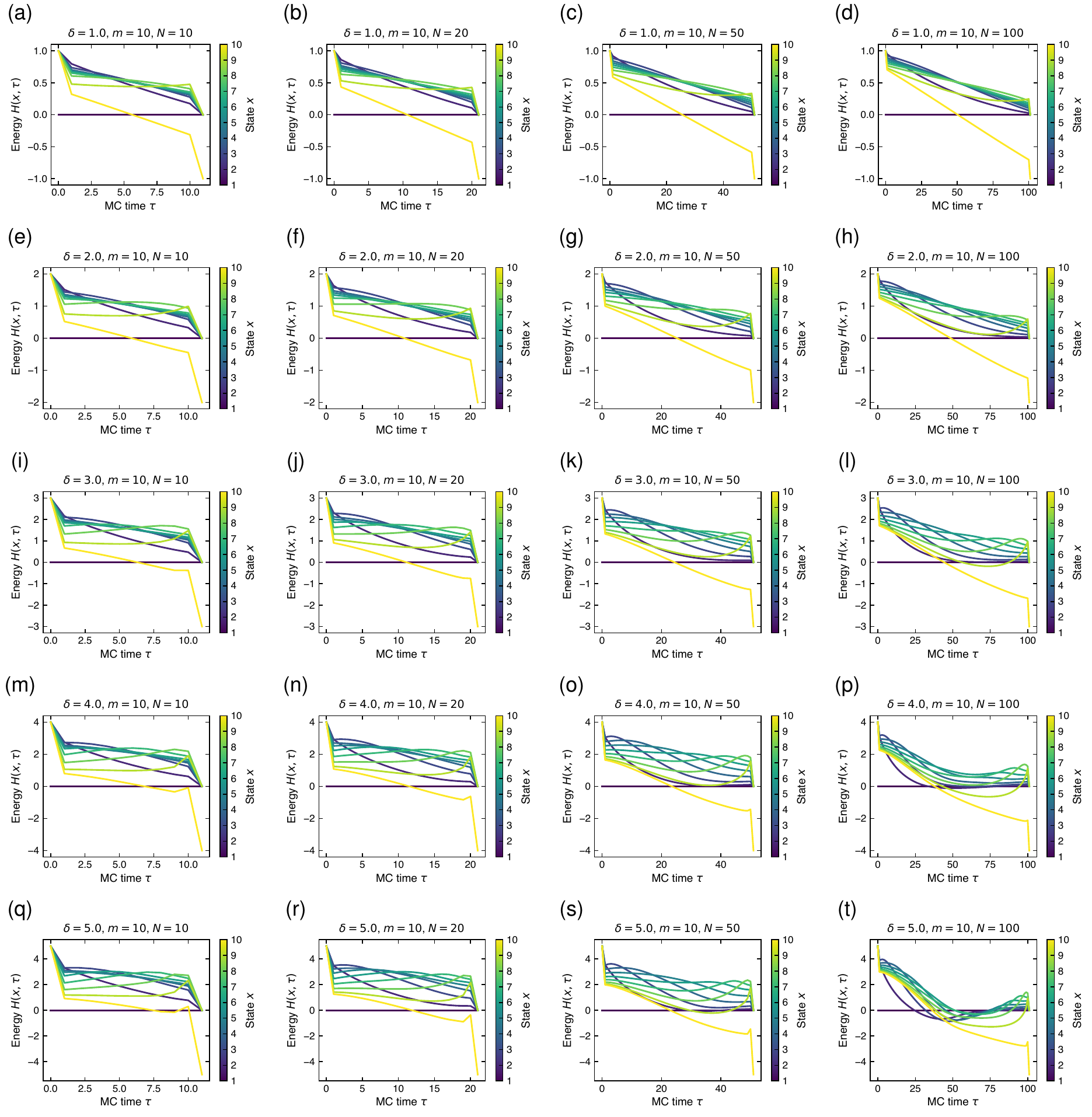}
\caption{\label{fig:exhaustive_protocols_pulling_single_well_8} 
Optimal intermediate Hamiltonians $H(x,\tau)$ of the shifted potential well with $m=10$ states and different depths $\delta$ and numbers of intermediates $N$.}
\end{figure*}

\begin{figure*}[ht]
\centering
\includegraphics[width=0.8\textwidth]{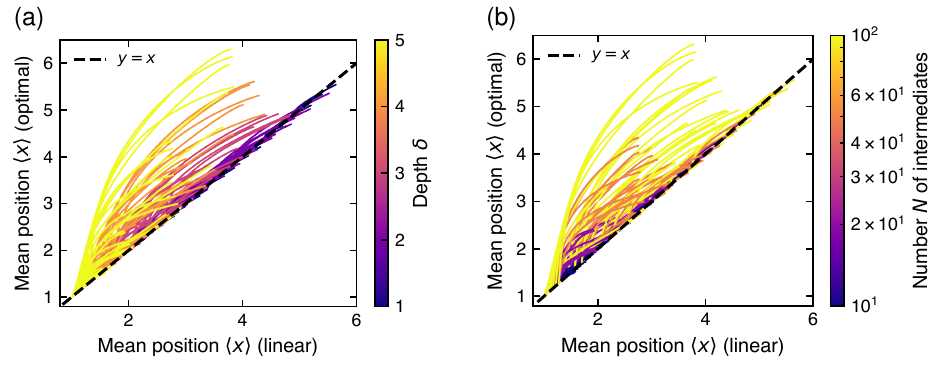}
\caption{\label{fig:correlation_plots_mean_position_esi} 
Correlation plots of the time-dependent mean position $\left\langle x\right\rangle$ obtained for optimal and linear intermediates.
The curves are colored according to the depth $\delta$ of the potential well.
(a): Correlation plot colored according to the depth $\delta$ of the potential well.
(b): Correlation plot colored according to the number $N$ of intermediates.}
\end{figure*}

\begin{figure*}[ht]
\centering
\includegraphics[width=0.8\textwidth]{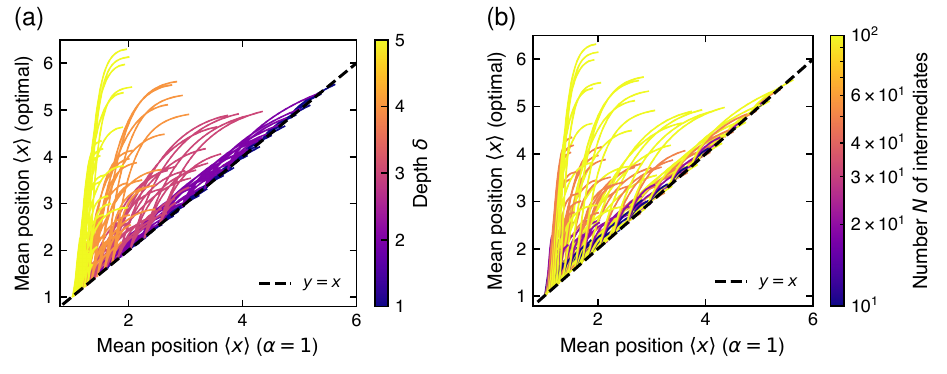}
\caption{\label{fig:correlation_plots_mean_position_log_linear_esi} 
Correlation plots of the time-dependent mean position $\left\langle x\right\rangle$ obtained for optimal and logarithmic ($\alpha=1$) intermediates.
The curves are colored according to the depth $\delta$ of the potential well.
(a): Correlation plot colored according to the depth $\delta$ of the potential well.
(b): Correlation plot colored according to the number $N$ of intermediates.}
\end{figure*}

\begin{figure*}[ht]
\centering
\includegraphics[width=0.8\textwidth]{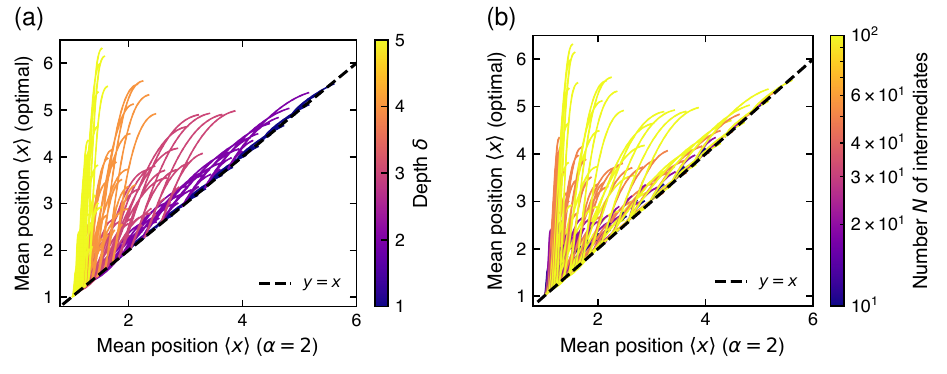}
\caption{\label{fig:correlation_plots_mean_position_avi_esi} 
Correlation plots of the time-dependent mean position $\left\langle x\right\rangle$ obtained for optimal and logarithmic ($\alpha=2$) intermediates.
The curves are colored according to the depth $\delta$ of the potential well.
(a): Correlation plot colored according to the depth $\delta$ of the potential well.
(b): Correlation plot colored according to the number $N$ of intermediates.}
\end{figure*}

\begin{figure*}[ht]
\centering
\includegraphics[width=0.8\textwidth]{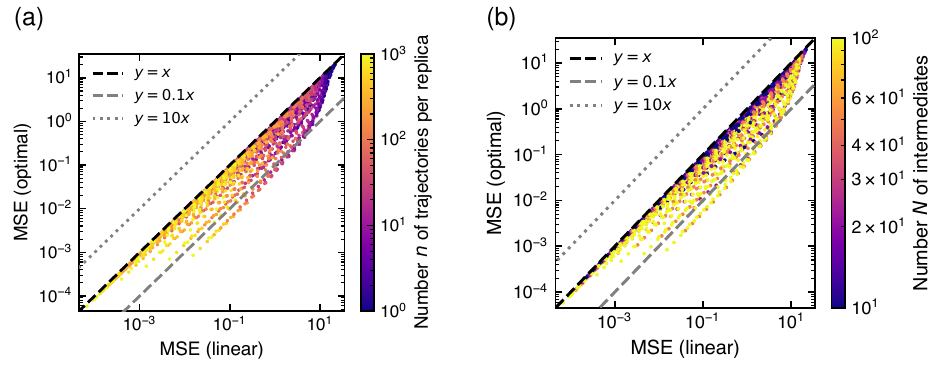}
\caption{\label{fig:correlation_plots_mse_single_well_esi} 
Correlation plots of the MSE for linear and optimal intermediates of the shifted potential well. 
Error bars corresponding to the standard error of the mean are included, but smaller than the symbol size.
(a): Correlation plot colored according to the number $n$ of trajectories per replica.
(b): Correlation plot colored according to the number $N$ of intermediates.}
\end{figure*}

\begin{figure*}[ht]
\centering
\includegraphics[width=\textwidth]{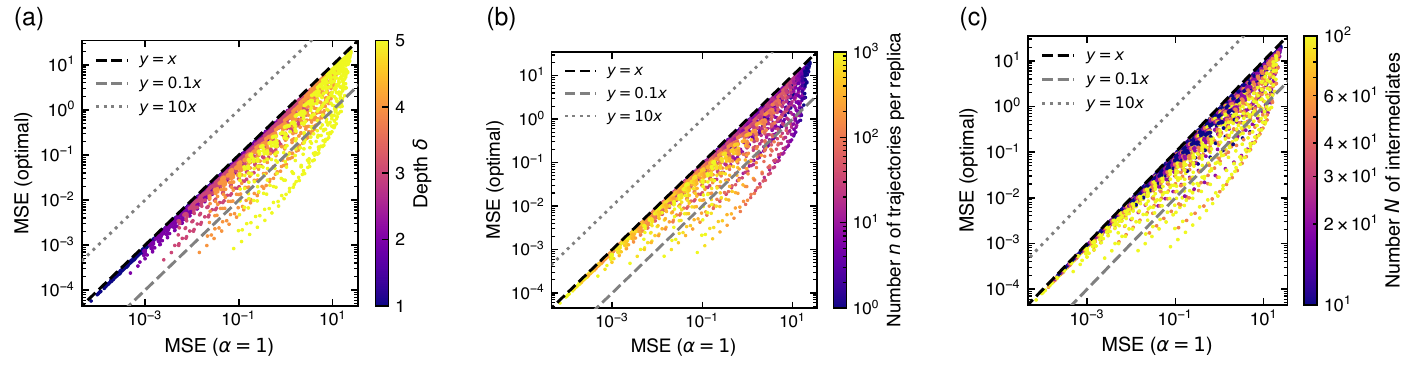}
\caption{\label{fig:correlation_plots_mse_single_well_log_linear_esi} 
Correlation plots of the MSE for logarithmic ($\alpha=1$) and optimal intermediates of the shifted potential well. 
Error bars corresponding to the standard error of the mean are included, but smaller than the symbol size.
(a): Correlation plot colored according to the depth $\delta$ of the potential well.
(b): Correlation plot colored according to the number $n$ of trajectories per replica.
(c): Correlation plot colored according to the number $N$ of intermediates.}
\end{figure*}

\begin{figure*}[ht]
\centering
\includegraphics[width=\textwidth]{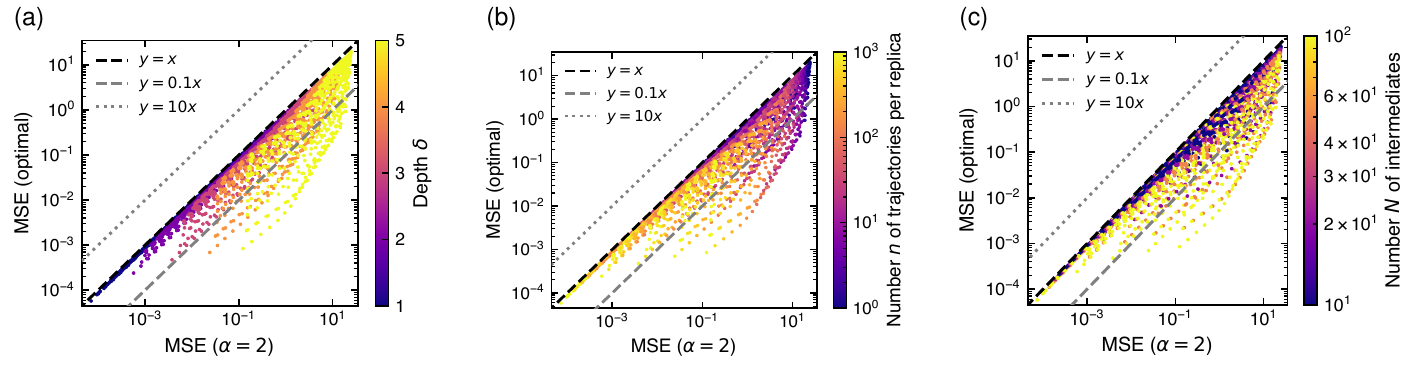}
\caption{\label{fig:correlation_plots_mse_single_well_avi_esi} 
Correlation plots of the MSE for logarithmic ($\alpha=2$) and optimal intermediates of the shifted potential well. 
Error bars corresponding to the standard error of the mean are included, but smaller than the symbol size.
(a): Correlation plot colored according to the depth $\delta$ of the potential well.
(b): Correlation plot colored according to the number $n$ of trajectories per replica.
(c): Correlation plot colored according to the number $N$ of intermediates.}
\end{figure*}

\begin{figure*}[ht]
\centering
\includegraphics[width=0.8\textwidth]{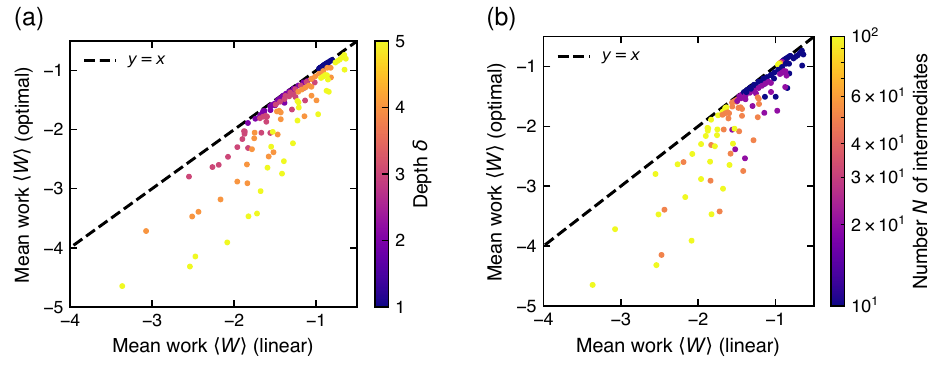}
\caption{\label{fig:correlation_plots_work_single_well_esi} 
Correlation plots of the mean work for linear and optimal intermediates of the shifted potential well.
Error bars corresponding to the standard error of the mean are included, but smaller than the symbol size.
(a): Correlation plot colored according to the depth $\delta$ of the potential well. 
(b): Correlation plot colored according to the number $n$ of trajectories per replica.}
\end{figure*}

\begin{figure*}[ht]
\centering
\includegraphics[width=0.8\textwidth]{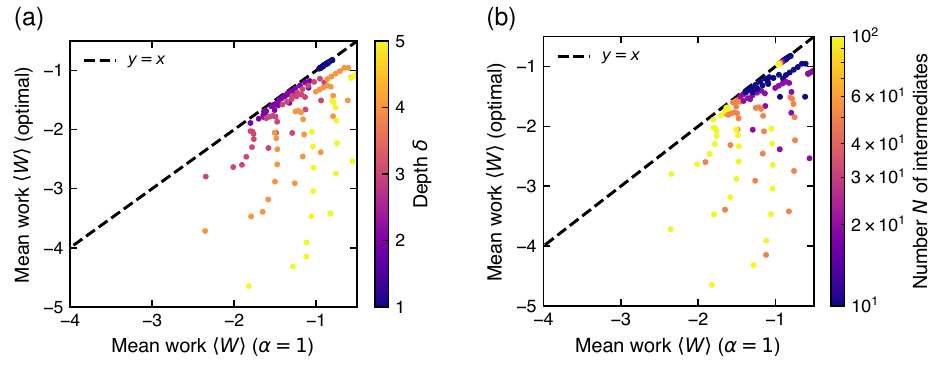}
\caption{\label{fig:correlation_plots_work_single_well_log_linear_esi} 
Correlation plots of the mean work for logarithmic ($\alpha=1$) and optimal intermediates of the shifted potential well.
Error bars corresponding to the standard error of the mean are included, but smaller than the symbol size.
(a): Correlation plot colored according to the depth $\delta$ of the potential well. 
(b): Correlation plot colored according to the number $n$ of trajectories per replica.}
\end{figure*}

\begin{figure*}[ht]
\centering
\includegraphics[width=0.8\textwidth]{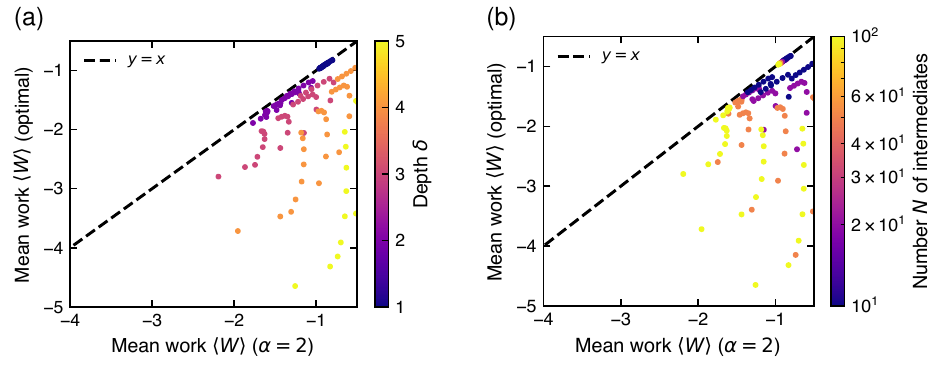}
\caption{\label{fig:correlation_plots_work_single_well_avi_esi} 
Correlation plots of the mean work for logarithmic ($\alpha=2$) and optimal intermediates of the shifted potential well.
Error bars corresponding to the standard error of the mean are included, but smaller than the symbol size.
(a): Correlation plot colored according to the depth $\delta$ of the potential well. 
(b): Correlation plot colored according to the number $n$ of trajectories per replica.}
\end{figure*}

\begin{itemize}
    \item Plots of the optimal intermediate Hamiltonians for the shifted potential well (\autoref{fig:exhaustive_protocols_pulling_single_well_1}--\autoref{fig:exhaustive_protocols_pulling_single_well_8})
    \item Correlation plots of the time-dependent mean position $\left\langle x\right\rangle$ obtained for optimal and linear intermediates of the shifted potential well (\autoref{fig:correlation_plots_mean_position_esi})
    \item Correlation plots of the time-dependent mean position $\left\langle x\right\rangle$ obtained for optimal and logarithmic ($\alpha=1$) intermediates of the shifted potential well (\autoref{fig:correlation_plots_mean_position_log_linear_esi})
    \item Correlation plots of the time-dependent mean position $\left\langle x\right\rangle$ obtained for optimal and logarithmic ($\alpha=2$) intermediates of the shifted potential well (\autoref{fig:correlation_plots_mean_position_avi_esi})
    \item Correlation plots of the MSE for linear and optimal intermediates of the shifted potential well (\autoref{fig:correlation_plots_mse_single_well_esi})
    \item Correlation plots of the MSE for logarithmic ($\alpha=1$) and optimal intermediates of the shifted potential well (\autoref{fig:correlation_plots_mse_single_well_log_linear_esi})
    \item Correlation plots of the MSE for logarithmic ($\alpha=2$) and optimal intermediates of the shifted potential well (\autoref{fig:correlation_plots_mse_single_well_avi_esi})
    \item Correlation plots of the mean work for linear and optimal intermediates of the shifted potential well (\autoref{fig:correlation_plots_work_single_well_esi})
    \item Correlation plots of the mean work for logarithmic ($\alpha=1$) and optimal intermediates of the shifted potential well (\autoref{fig:correlation_plots_work_single_well_log_linear_esi})
    \item Correlation plots of the mean work for logarithmic ($\alpha=2$) and optimal intermediates of the shifted potential well (\autoref{fig:correlation_plots_work_single_well_avi_esi})
\end{itemize}

\newpage

% Double-Well

% Shifted Well

